\documentclass[fleqn,usenatbib]{mnras}
\usepackage[T1]{fontenc}
\usepackage{ae,aecompl}
\usepackage{graphicx}	
\usepackage{amsmath}	
\usepackage{amssymb}
\usepackage{pdflscape}
\usepackage{siunitx}
\usepackage{dcolumn}
\usepackage{multicol} 
\usepackage{hyperref}
\defcitealias{2022MNRAS.511.4882S}{paper I}

\title[UV luminosity function at $z \sim 0.6-1.2$]{{On the bright-end of the UV luminosity functions of galaxies at $z \sim 0.6-1.2$}}

\author[M. Sharma et al.]
       {{M. Sharma$^{1,2}$\thanks{E-mail: mnushv@gmail.com (MS)},
           M. J. Page$^{1}$,
           I. Ferreras$^{3,4,5}$,
           A. A. Breeveld$^{1}$}\\
         $^1$Mullard Space Science Laboratory,
         University College London,
         Holmbury St Mary, Dorking, Surrey, RH5 6NT, UK\\
         $^2$Isaac Newton Group of Telescopes,
         C. Álvarez Abreu, 70, E38700 Santa Cruz de La Palma, 
         La Palma, Spain\\
         $^3$Instituto de Astrofísica de Canarias, 
         Calle Vía Láctea s/n, E38205, 
         La Laguna, Tenerife, Spain\\
         $^4$Departamento de Astrofísica, 
         Universidad de La Laguna, E38206, 
         La Laguna, Tenerife, Spain\\
         $^5$Department of Physics and Astronomy, 
         University College London, 
         Gower Street, London WC1E 6BT, UK}
         
\date{Accepted 2024 May 14; Received 2024 May 7; in original form 2023 January 5}
       
\pubyear{2024}

\begin{document}
\label{firstpage}
\pagerange{\pageref{firstpage}--\pageref{lastpage}}
\maketitle

\begin{abstract}
We derive the Ultra-Violet (UV) luminosity function (LF) of star-forming galaxies in the redshift
range $z = 0.6 - 1.2$, in the rest-frame far-UV (1500 {\AA}) wavelength.
For this work, we are in particular interested in the bright end of the UV LF in 
this redshift range.
Data from the \textit{XMM-Newton} Optical Monitor
(XMM-OM), near-ultraviolet (2410$-$3565 {\AA}) observations over 1.5 deg\textsuperscript{2} of the COSMOS field are employed for this purpose. 
We compile a source-list of 879 sources with $UVW1_\mathrm{AB}$ in the range 
$\sim 21-24$ mags from the wide-area UVW1 image of the COSMOS field
in the two bins $0.6 \leq z \leq 0.8$ and $0.8 \leq z \leq 1.2$.
The $M_{1500}$ for these sources lies in the interval $[-19.10,-22.50]$.
We use the maximum likelihood to fit a Schechter function model to the un-binned
data to estimate the parameters (faint-end slope, characteristic magnitude,
and normalisation) of the Schechter function.
We find the shape of the LF to be consistent with the Schechter model, and
the parameters are in fair agreement with other studies conducted using
direct measurements of the 1500 {\AA} flux.
We see a brightening of the characteristic magnitude as we move from
lower (0.7) to higher (1.0) redshift.
The measures for luminosity density are within the error margins 
of past studies.
We examine the brightest sources in our sample for the AGN contribution. These
sources are characterised by their spectral energy distributions,
integrated infrared luminosities, and morphologies.
We also explore their overlap with the brightest IR galaxies in a similar redshift 
range.

\end{abstract}

\begin{keywords}
galaxies: evolution - ultraviolet: galaxies - ultraviolet: luminosity function - galaxies: luminosity function 
\end{keywords}

\section{Introduction}
\label{sec:1}

The luminosity function (LF) is one of the most powerful statistical tools employed
to study the distribution of large-scale properties (e.g. galaxy masses, luminosities, star formation rates) of a galaxy population under consideration.
Galaxies can be arranged according to how much light is coming out of them. 
Counting these galaxies in a luminosity bin per comoving
volume gives a number density traditionally called the luminosity function.
In other words, a luminosity function describes the number 
density of galaxies in a luminosity interval.

The estimation of galaxy LFs remains an important task in 
extra-galactic astronomy and observational cosmology, for it has 
multiple applications.
The integral of the luminosity function weighted by luminosity can be used 
to derive the luminosity density, which scales directly
with the contribution of the galaxy population under consideration to the star-formation rate density \citep[e.g.][]{1996ApJ...460L...1L,1998ApJ...498..106M}.
The LF is widely used to de-project the angular correlation function 
using Limber’s Equation \citep{1953ApJ...117..134L} to estimate
the three-dimensional spatial distribution of galaxies, and to study the 
large-scale properties like the correlation length, halo mass, galaxy 
bias, and halo occupation \citep[e.g.][]{2009A&A...498..725H}.
There are studies proposing to constrain the primordial non-Gaussianities at small scales \citep{2021JCAP...01..010S}, and to study the primordial
power spectrum \citep{PhysRevD.102.083515}, using the UV LFs at high
redshifts.
A precise estimate of the faint-end slope is required to estimate the
magnification bias in gravitational lensing studies \citep[e.g.][]{1993LIACo..31..217N}, which can be used as an independent
probe to study cosmological parameters such as the total
matter density, the dark matter power spectrum normalisation and also
the galaxy bias \citep[e.g.][]{Scranton_2005,2009A&A...507..683H}.

The theoretically predicted shape from the $\Lambda-$CDM model comes in
the form of the dark-matter halo mass function \citep[e.g.][]{2021ApJ...922...89S}. Assuming a baryon
fraction and star formation efficiency, the halo mass function can be compared to an observed stellar mass function
\citep{2017ARA&A..55..343B}. The comparison leads to
a mismatch between the two, which becomes severe at the two extremes
\citep[high/low stellar masses or bright/faint luminosities;][]{2001MNRAS.326..255C,2003MNRAS.339.1057Y,2010ApJ...717..379B}. Due
to this mismatch, there is a lot of work going on to understand
the nature of physical processes that might be causing the deviations
in the observed LF shape with respect to the theoretical predictions.
This makes 
accurate estimates of the LF even more important as these can 
provide important insights into the additional physical processes 
occurring at the small scales and help us understand the baryon cycle 
at those scales \citep[][]{2015ARA&A..53...51S}.
Variation in the shape of the LF as a function of redshift can be used 
as a proxy for the evolution of galaxies between different epochs in the
history of the Universe.

The stellar evolution in a normal galaxy is believed to dictate many
radiative processes occurring in the galaxy and control the amount of
light produced by it. The dust in the galaxy controls the fraction of this UV
light coming out of the galaxy. The
UV luminosity of galaxies is of particular interest, as it is produced
mainly by massive stars with short lifetimes \citep[][]{2014ARA&A..52..415M}, and can be used to
trace the underlying star formation activity in galaxies over a timescale of
100 Myr. The rest-frame 1500 {\AA} emission in the UV is
used extensively in the literature, as it is one of the most important tracers for understanding the unobscured star formation in galaxies.
\citep[][]{2012ARA&A..50..531K}.

Using both ground-based and space-borne observatories, many studies have 
calculated the UVLF at redshift $> 1.2$
\citep[e.g.][]{2006ApJ...653..988Y,2010ApJ...720.1708H,2016MNRAS.456.3194P,
            2020MNRAS.494.1894M}.
However, in the redshift range $z < 1.2$, so far only a handful of studies have been carried out because the emission of 1500 {\AA} can only be accessed
from space-borne instruments. The first results on the galaxy UV LF in
the redshift range $\left(0.2 \leq z \leq 1.2\right)$ were obtained by
\citet{2005ApJ...619L..43A} using NUV data from the \textit{Galaxy Evolution 
Explorer satellite} \citep[GALEX;][]{2005ApJ...619L...1M}.
These results were followed by \citet{2010ApJ...725L.150O}
who used data from the \textit{Hubble Space Telescope (HST)} WFC3/UVIS instrument
to explore the redshift range $0.5 < z < 1$, and by \citet{2015ApJ...808..178H} who used the UV/Optical Telescope
(UVOT) on the \textit{Neil Gehrels Swift} observatory to calculate the LF in the
redshift region from 0.2 to 1.2.
Using ground-based observations from the CFHT and Subaru Telescope, \citet{2020MNRAS.494.1894M} and from VLT \citet{refId0} calculated
the Galaxy LF. \citet{2020MNRAS.494.1894M} have re-analysed the GALEX
data from \citet{2005ApJ...619L..43A} to extend their luminosity function
to redshifts less than 0.9.
Very recently \citet{2021MNRAS.506..473P} have published their UV LF
results in the redshift range of $0.6 < z < 1.2$ using the observations 
of the 13 Hr field taken through the UVW1 filter of the Optical Monitor telescope onboard the \textit{XMM-Newton} observatory \citep[XMM-OM, Optical Monitor;][]{2001A&A...365L..36M}.
In \citet{2022MNRAS.511.4882S}, hereafter \citetalias{2022MNRAS.511.4882S}, we used data from
observations of the Chandra Deep Field South (CDFS) taken by the same instrument (i.e. XMM-OM) to estimate the LF from redshift 0.6 to 1.2.
This current work is a follow-up study to \citetalias{2022MNRAS.511.4882S} using the UVW1 imaging obtained with XMM-OM in the wide area Cosmic evolution survey field \citep[COSMOS;][]{2007ApJS..172....1S}. With the wide area of this 
field, we expect to find many luminous sources in our survey to extend our LF
to brighter absolute magnitudes.
The UVW1 filter ($\lambda_{\mathrm{cen}} = $ 2900 {\AA}) is close to the rest-frame 1500 {\AA} wavelength region in the redshift range of 0.6 to 1.2.
The full width at half-maximum (FWHM) of the XMM-OM UVW1 point spread function (PSF) is $\simeq 2$ arcsec\footnote{\url{https://xmm-tools.cosmos.esa.int/external/xmm_user_support/documentation/uhb/XMM_UHB.html}}.
In comparison the \textit{GALEX} NUV filter falls bluewards of the rest-frame 1500 {\AA} in this redshift range. 
Moreover, the sharper PSF of the XMM-OM compared to \textit{GALEX}, which is around 5 arcsec \citep{2007ApJS..173..682M}, makes it easier to assign the correct optical counterparts to XMM-OM sources than to  \textit{GALEX} sources.

Dust in star-forming galaxies absorbs and scatters the ultraviolet (UV) photons produced by young stars. So, the amount of UV radiation coming out of the galaxies depends on the dust content of the galaxy. Fortunately, this obscured light is re-emitted in the
FIR. Since this work is mainly about the bright end of the LF, we take a look at the properties of the sources in the bright bins of the LF and try to establish a connection between the bright UV galaxies and their infrared (IR) counterparts.
We will undertake an in-depth study of the IR properties of galaxies in the CDFS and COSMOS
fields in future work.

The remainder of this paper is structured as follows. All data and
processing (that is, observations, UVW1 data, image processing,
and ancillary data used for redshift
information and to identify stars and active galactic nuclei)
are explained in Section \ref{sec:2}. 
The completeness simulations are also explained in this Section.
The corrections for Galactic extinction and the K-corrections are discussed in Section \ref{sec:3}, before the final analysis.
The methods used to compute the binned LF, fit the Schechter function parameters,
and estimate the luminosity density are described in Section \ref{sec:4}. 
This Section also includes a description of the expected effects of cosmic variance. 
In addition, we provide details for the fitting spectral
energy distributions to the sources in the brightest bins
of the LF in the same Section.
We present our results and discuss them in Section \ref{sec:6}. 
Finally, we conclude this paper in Section \ref{sec:7}.
For this paper, we have assumed a flat cosmology with
$\Omega_{\Lambda}=0.7$, $\Omega_{M}=0.3$ and Hubble's constant 
$H_0=70$\,km\,s$^{-1}$\,Mpc$^{-1}$. The distances are
calculated in comoving coordinates in Mpc. The AB system of magnitudes
\citep{1983ApJ...266..713O} is adopted throughout this study.

\section{Observations \& Data}
\label{sec:2}
\subsection{XMM-OM observations of the COSMOS field}
\label{sec:2.1}

\begin{figure*}
  \hspace*{-0.1cm}\centering
  \includegraphics [width=0.95\textwidth]{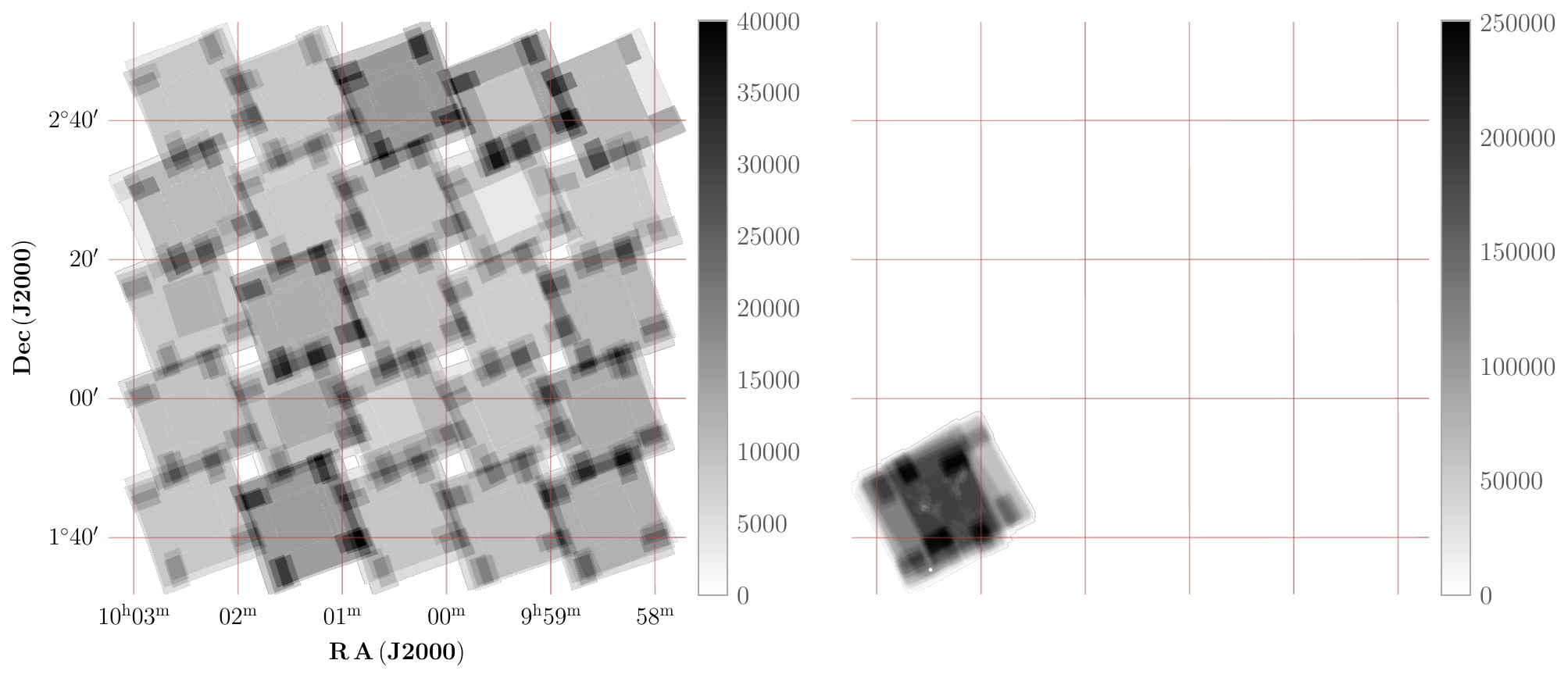}
  \caption{COSMOS vs. CDFS comparison. \textit{Left}: We show the exposure map of the COSMOS UVW1 image as a $5 \times 5$ grid of 25 pointings (subfields). 
  The small squares/rectangles inside the COSMOS image indicate gaps between the different pointings, where there is no OM data. The colour bar represents the exposure time scale in seconds. The exposure time per pixel and the sky-area of the final COSMOS mosaics are 13 Ks
  and 1.56 sq. deg.
  \textit{Right}: Exposure map of the CDFS UVW1 image on the same scale as the COSMOS image. It can be seen that, compared to the CDFS image, the COSMOS image has a much larger sky area, but a shallow depth. The exposure times per pixel for the final CDFS image are 130 Ks and 396 sq. arcmins.
  Note that the colour bars for both images have different scales.}
  \label{fig:uvw1_im}
\end{figure*}

The COSMOS field
with its wide 2 deg\textsuperscript{2} survey area and multi-wavelength
coverage from X-ray to radio has been catering for studies of galaxy evolution and large-scale structure over a wide range of redshifts.
In this study, we use data from XMM-COSMOS \citep{2007ApJS..172...29H,2007ApJS..172..341C}, a wide-field survey of the COSMOS field with the XMM-Newton observatory.
In particular, the data taken by the UVW1 filter (2410$-$3565)\footnote{\url{https://www.mssl.ucl.ac.uk/www_astro/XMM-OM-SUSS/SourcePropertiesFilters.shtml}} {\AA}
of XMM-OM at a central wavelength of 2910 {\AA} are used. The UVW1 data are most relevant to
derive the 1500 {\AA} LF at the redshifts $0.6 - 1.2$, as it minimises the range of
K-corrections made to the UVW1 magnitudes.
For this filter, XMM-OM provides a PSF (FWHM) of 2 arcsec.
Our catalogue of the XMM COSMOS Survey consists of images from
55 observations between December 2003 and May 2007, targeting 25 slightly overlapping pointings spanning the COSMOS field. These 25 pointings are
arranged in a $5\times5$ grid. The final image also contains a $4\times4$ grid of holes (without data) with typical sizes ranging from 4 to 7.5 arcmin$^2$ 
(Fig. \ref{fig:uvw1_im}).

All the imaging is produced in the
`default' (also known as `rudi-5') configuration, where a setup of 5 consecutive 
exposures is used to cover 92 percent of the $17\times17$ arcmin$^2$
FOV of the XMM-OM. For details of the image configurations, see
\citet{2001A&A...365L..36M}. More details about the
different image configurations can also be found in the XMM-Newton User's Handbook\footnote{\url{https://xmm-tools.cosmos.esa.int/external/xmm_user_support/documentation/uhb/omdef.html}}.

\subsection{XMM-OM data reduction and processing}
\label{sec:2.2}

The processing of COSMOS data is done here following the same process
as in \citetalias{2022MNRAS.511.4882S} (i.e. XMM-OM image processing pipeline with
some tweaks). We give a summary of the process here and refer the reader to \citetalias{2022MNRAS.511.4882S} and \citet{2012MNRAS.426..903P,2021MNRAS.506..473P} for
details.
The raw data were obtained from the XMM-Newton Science Archive\footnote{\url{https://www.cosmos.esa.int/web/xmm-newton/xsa/}}
(XSA). 
The standard XMM-Newton Science Analysis System\footnote{\url{https://www.cosmos.esa.int/web/xmm-newton/sas}}
task \textsc{omichain} is used for the primary pipeline processing of
the data.
The data products are removed from the pipeline after correcting for
mod-8 pattern for additional processing
to get rid of the background scattered light feature at the centre of the images, 
and also to remove some artefacts. The artefacts and their 
corrections are explained in Section 2.2 of \citetalias{2022MNRAS.511.4882S}.
After correcting for all the cosmetic effects, the images were then distortion
corrected for any nonlinear offsets in pixel positions, aligned with the
equatorial coordinate frame of SDSS DR12 \citep{2015ApJS..219...12A},
then rotated and re-binned into sky coordinates using the SAS task
\textsc{omatt}.
Finally, the images were co-added using the SAS task \textsc{ommosaic}. 
The spatial extent of the final co-added COSMOS image (1.56 sq. deg.) can be seen in 
Fig. \ref{fig:uvw1_im} as compared to the smaller but deeper CDFS 
image used in the \citetalias{2022MNRAS.511.4882S}.

The final image is then fed to the SAS task \textsc{omdetect}, which performs 
the photometry and source detection, using different algorithms for
identifying points and extended sources. The details of \textsc{omdetect} can be found in \citet{2012MNRAS.426..903P}, and we summarise the appropriate details in the text that follows.
For point-like sources, the default photometric aperture used by {\sc omdetect} depends on the signal-to-noise ratio (in particular the ratio of the source counts and the rms background fluctuation ratio within an aperture of the same size), with brighter sources measured in a 5.7 arcsec radius aperture, and fainter sources measured in a 2.8 arcsec aperture. Most of the UVW1-detected galaxies with redshifts between 0.6 and 1.2 appear point-like to XMM-OM, and the brightest of these would by default be measured in a 5.7 arcsec radius aperture, unless there are one or more neighbours close to the source. In this case, the aperture size is reduced depending on the distance to the nearest neighbour. The minimum radius of the aperture is set to 2.8 arcsec. Inspection of deep optical images of the luminous galaxies in UVW1 image reveals that photometry in such a large aperture will be contaminated by the light from other surrounding galaxies. Therefore we adopted the smaller (2.8 arcseconds radius) aperture for all the galaxies within our sample. 
The magnitudes of sources with reduced apertures are adjusted to their equivalent value in a 5.7 arcsec radius aperture by applying an aperture correction. This involves extrapolating the counts using a curve of growth, which is based on the calibrated point spread function (PSF) specific to the corresponding bandpass.
In total, $7027$ sources are detected in the UVW1 image above a signal-to-noise threshold of 4.
The edges of the images are areas of low signal-to-noise, resulting in spurious sources entering the catalogue. To eliminate this problem, we mask the outer 10 pixels of the COSMOS UVW1 image. Due to this masking,
we lost 0.046 deg$^2$ (2.9 percent) of sky area and 102 (1.4 percent) of the sources.

\subsection{Completeness}
\label{sec:2.3}

\begin{figure}
  \hspace*{-0.1cm}\includegraphics[width=0.95\columnwidth]{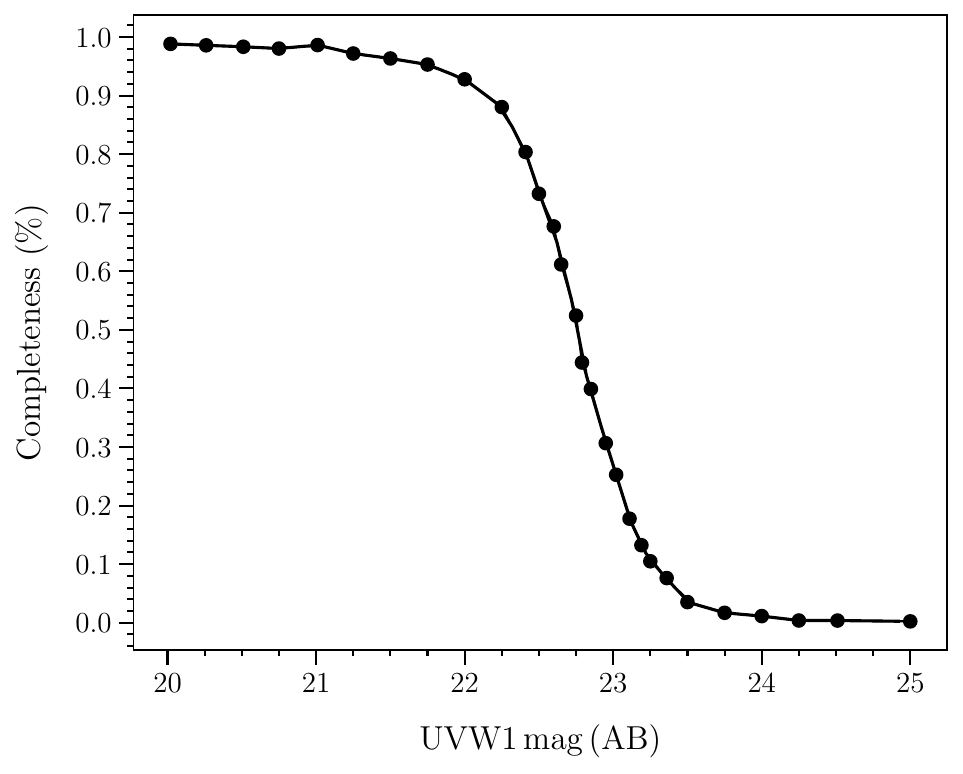}
  \caption{Completeness of the source detection as a function of UVW1
  magnitude, as determined from the simulations described in Section 
  \ref{sec:3.1}. The black data points represent a fraction of recovered
  simulated galaxies at each input UVW1 mag.}
  \label{fig:comp}
\end{figure}

\begin{figure}
  \hspace*{-0.1cm}\includegraphics[width=1.0\columnwidth]{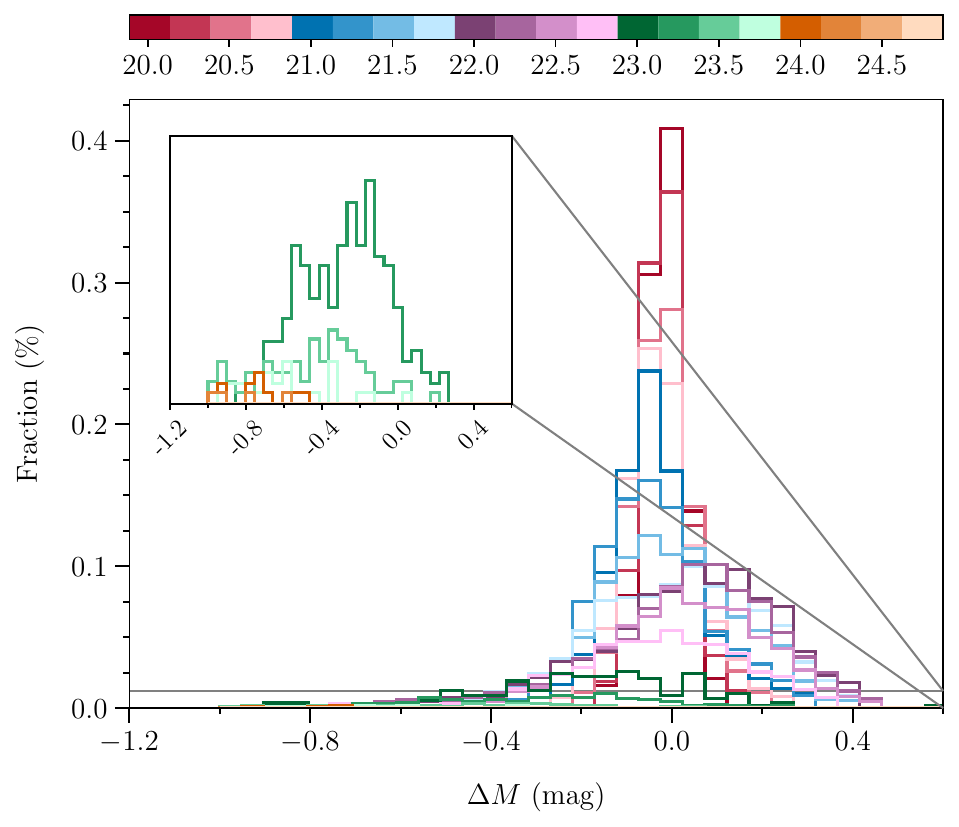}
  \caption{The distributions of the magnitude errors in the detection process at each simulated UVW1 magnitude, where the x-axis shows the
  magnitude offsets between the inserted and recovered sources and the
  the y-axis represents the fraction of sources at each offset. Each colour, 
  as represented by the discrete colour bar represents the input 
  magnitude in the completeness simulations.
  The inset window represents a part of the plot stretched in the y-axis 
  for clarity.}
  \label{fig:hist}
\end{figure}

\begin{figure}
  \hspace*{-0.1cm}\includegraphics[width=0.95\columnwidth]{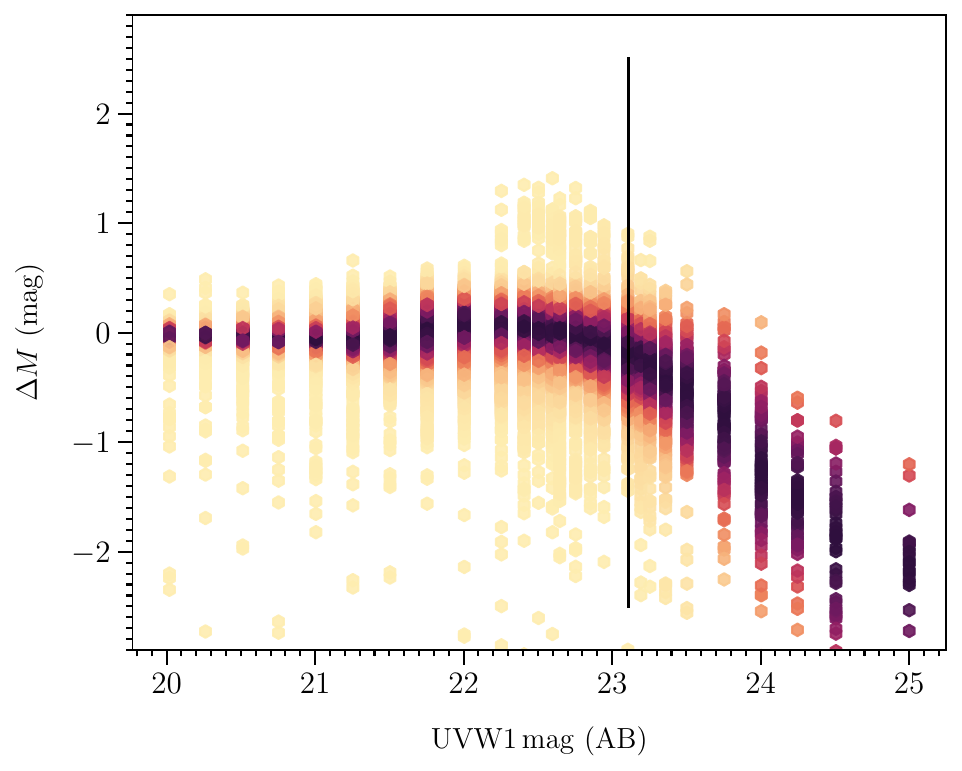}
  \caption{The magnitude errors at each simulated input UVW1 magnitude. The colour map from light to dark colours represents the increasing fraction of recovered sources at each input magnitude. The black line shows the magnitude limit of our survey beyond which no sources are included in the further analysis.}
  \label{fig:hist2}
\end{figure}

The completeness of a survey can be defined as the probability of detecting a source with a particular magnitude. This probability is controlled by various factors, primarily signal-to-noise ratio, but also the
blending of two or more faint sources into one, i.e. source confusion. 
To quantify how complete (or incomplete), as a function of
magnitude, our galaxy sample is, 
we use the same technique as in \citetalias{2022MNRAS.511.4882S}. We briefly discuss the process here
and refer the reader to \citetalias{2022MNRAS.511.4882S} and \citet{2021MNRAS.506..473P} for details.

We simulate artificial point
sources of varying magnitudes, add them to the real image one at a time, and detect them using \textsc{omdetect} (Section \ref{sec:2.2}). If a source
is detected within a circular region of radius 1 arcsec of the position
of the inserted source, we consider the
source to be recovered. 
The simulation effectively incorporates blending effects and accurately accounts for the impact of the Eddington bias and other factors. 
Its accuracy is ensured by adding sources one at a time to the COSMOS UVW1 image.
The stepwise addition of each source, also
ensures that the UV number count is not perturbed and each inserted source will be affected by other sources in the image (almost) exactly as a source at that flux.
At least 1000 sources are used at each input magnitude ranging from
UVW1 mag 20 to 25 in steps of 0.25 mag, with smaller step sizes of 0.05, 0.10 and 0.15
between magnitudes 22.25 to 23.50, where completeness drops quickly. 
Before we calculate the quantities required for our further analysis,
the simulations are corrected for the reddening and the edge-mask applied to our final image (Section \ref{sec:2.2}).

The recovery fraction at a particular magnitude, which is also the probability of detecting a source at that magnitude, defines completeness. 
This fraction is plotted as a function of UVW1 AB
magnitude in Fig. \ref{fig:comp} and is used in Section 
\ref{sec:4.1} to calculate the binned LF.
The distributions of the magnitude offsets between the inserted and 
recovered sources are used as the error distributions, as they contain
information of the unknown systematics unique to the XMM-OM, the 
UVW1 filter and the detection process.
These distributions are plotted as a function of the simulated UVW1
magnitudes in Fig. \ref{fig:hist}. Each colour represents a simulated
UVW1 magnitude ranging from UVW1 mag 20 to 24.75.
The scatter observed in these distributions arises from the statistical errors associated with the measurements of the magnitudes at each input magnitude. The bias accounts for the variations caused by fluctuations in the background estimate and other systematic factors. 
These perturbations (both scatter and bias) in the apparent magnitudes translate directly to perturbations in the absolute magnitudes. 
The distributions of magnitude offsets are forward-folded with the LF models to obtain models for the observed distributions, before the maximum likelihood fitting is performed (Section \ref{sec:4.2}) to obtain the LF parameters \citep[also see][]{2021MNRAS.506..473P,2022MNRAS.511.4882S}.
We note here that no corrections are made to the observed magnitudes, and that the error distributions are only used in the maximum likelihood fitting, not in the binned luminosity functions.

As can be seen in Fig. \ref{fig:comp}, our catalogue is found to be 
> 98 percent complete for
UVW1 < 21 mag, 75 percent complete at UVW1 $\leq$ 22.5 mag and falls 
< 10 percent as UVW1 magnitude goes beyond $23.25$. 
At UVW1 24 mag the recovery fraction decreases to $< 1$ percent. 
From Fig. \ref{fig:hist} (see inset), it can be seen that the magnitude offset distribution is composed predominantly of negative values for UVW1 mag 23.75 and completely for UVW1 mag > 24, implying that the recovered sources are brighter than the ones inserted.

This means that sources fainter than UVW1 mag
24 are only detected due to flux boosting - faint sources blended with the noise spikes. 
We notice that the faintest output magnitude of a detected source is 24.25, 
which is also apparent from the completeness plot, where above 24.25
magnitudes the completeness curve becomes more or less an asymptote
to the magnitude axis. 
We conclude that
at around this magnitude (UVW1 24.25 mag), we hit the sensitivity limit for
the survey (i.e. at fainter magnitudes than this we can not get any detections) and we are not limited by confusion.
To be on the safe side, we take the conservative approach 
similar to \citetalias{2022MNRAS.511.4882S} and 
apply a magnitude limit brighter than the faint detection limit. 
We chose UVW1 23.11 mag (= 23.02 mag after reddening correction) 
as the magnitude limit for our survey. The completeness level at this
magnitude is $47$ times the residual level at the sensitivity limit.

The distributions of magnitude offsets are plotted in Fig. \ref{fig:hist2}
as a function of the input magnitude. The colours go from light to dark as the recovery fraction increases at each input magnitude. Notably, the peak of the distributions (represented by the darkest colours) are predominantly centred around the zero offset, up until the black line denoting the magnitude limit of our survey.
However, the peaks of the distributions become flatter, and their centres deviate from the zero offset position, beyond the magnitude limit of the survey. This observation serves as a reassuring sanity check, validating the reliability of the magnitude limit implemented in our survey.

\subsection{Ancillary data}
\label{sec:2.4}

\begin{figure}
  \centering
  \hspace*{-0.1cm}\includegraphics [width=0.95\columnwidth]{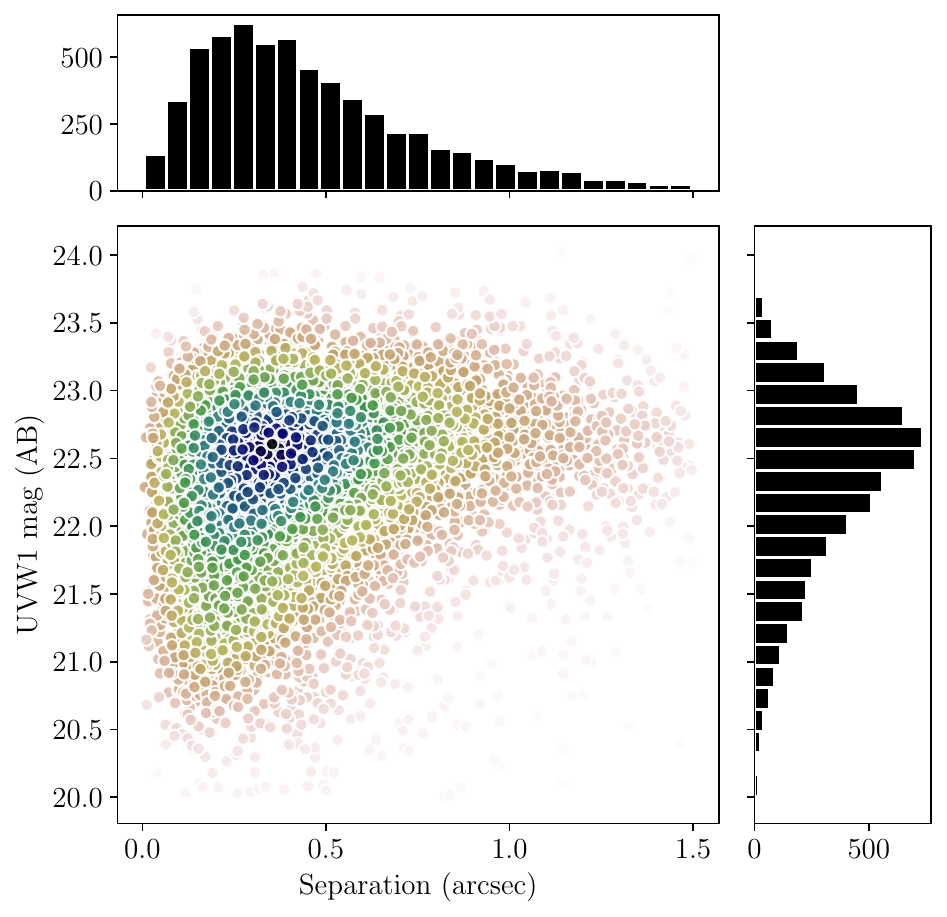}
  \caption{A joint distribution of the UVW1 AB magnitudes of the sources
   and the angular offsets with their respective counterparts after 
   matching them with the COSMOS2015 catalogue using 1.5 arcsec as the maximum offset radius. The histograms represent the marginal distributions.}
  \label{fig:mag_offset_dis}
\end{figure}

Once we have our source list, with UVW1 magnitudes, we combine it
with other catalogues from the literature to collect additional
information about our sources.
We use this additional information to identify the stars and AGNs
in our sample, and to assign redshifts to our galaxies.
Similarly to \citetalias{2022MNRAS.511.4882S}, we perform a test to find out the appropriate
search radius to cross-correlate our source list with other catalogues.
As we are matching a large catalogue of UVW1 sources to a very deep multi-wavelength catalogue with a much higher sky density, a naive positional match has the potential to generate a large number of spurious matches. Therefore, we have carefully explored the matching criteria.

There should be a limit to how blue the UVW1$-u$ colours of star-forming 
galaxies can be. Therefore, a UVW1$-u$ colour-cut can be used with a large matching radius to
achieve a high matching rate while keeping spurious matches to a minimum.
We find this blue limit in Appendix~\ref{sec:cross-corr},
by comparing the UVW1 and $u$ colours of stars, QSOs and galaxies. 
We also check what fraction of total sources are spuriously matched, in catalogues compiled using different matching radii.
As described in Appendix~\ref{sec:cross-corr}, we obtain a blue limit of UVW1$-u = -0.5$.
Taking into account the uncertainties on the magnitudes, we put a conservative limit
UVW1$-u>-1$ on our sources and find a matching radius of 1.5 arcsec. 

The distribution of the UVW1 magnitudes of the sources and their distances from counter-parts in the COSMOS2015 catalogue are plotted in Fig. \ref{fig:mag_offset_dis}. We caution the reader here that the angular offsets reflect the scatter in position differences between the UVW1 positions and the COSMOS2015 positions rather than a systematic offset.

\subsubsection{Stars}
\label{sec:2.4.1}
We have used several different sources to identify the stars in our
source list. The first is the catalogue from
\citet{2007ApJS..172..219L} which uses two alternate methods to
differentiate the stars from other sources.
We also use spectroscopic identifications of stars from
\citet{2006ApJ...644..100P,2009ApJ...696.1195T}; SDSS - III \citep{2015ApJS..219...12A} and \citet{2018ApJ...858...77H}.

In addition, we employ GAIA DR2 data \citep[][]{2018A&A...616A...1G} and
remove sources with significant proper motions. We define the significant
proper motions to be those for which the values are at least more than three
times the standard errors.
We also use \citet{2016ApJ...817...34M}, which uses both spectroscopic identification and photometric SEDs to identify stars.
In total, we identified 1523 stars in our catalogue,
which constitute $\sim$ 23 percent of sources in our edge-masked COSMOS UVW1 source list. For the rest of the analysis, we remove these stars from
our source list.

\subsubsection{Redshifts}
\label{sec:2.4.2}

We merge the remainder of our sources with catalogues containing
spectroscopic and photometric redshifts. Following are the brief
details of the catalogues.\\
We use redshifts from the hCOSMOS \citep[][]{2018ApJS..234...21D}, which
surveyed the COSMOS field with 
the MMT Hectospec multi-fibre spectrograph. We prioritise their data from
the central $\sim$ 1 deg$^2$ of the field, which
(hCOS20.6) is 90 percent complete to the limiting $r$ magnitude of 20.6.
We used a set of spectroscopic redshifts compiled by \citet{2021MNRAS.501.6103A} released as part of their PAUS photometric
catalogue. Then we add data from the zCOSMOS-bright survey \citep{2007ApJS..172...70L,2012ApJ...753..121K}, which provides
high-quality spectroscopic redshifts up to redshift 1.2 for sources
with $I_\mathrm{AB} \leq 22.5$.
Next, we use the catalogue from 
\citet{2018ApJ...858...77H}, who used the Deep Imaging Multi-Object 
Spectrograph (DEIMOS) on the Keck II telescope. 
We further add spectroscopic redshifts from the following: Fibre Multi-Object Spectrograph (FMOS) - COSMOS survey \citep{2019ApJS..241...10K} which uses
medium resolution spectroscopy to obtain > 5000 redshifts,
the PRism MUlti-object Survey \citep[PRIMUS;][]{2011ApJ...741....8C} targeting
faint galaxies with the Inamori Magellan Areal
Camera and Spectrograph (IMACS) spectrograph on the Magellan I Baade 6.5 m telescope and
the Large Early Galaxy Census \citep[LEGA-C;][]{2016ApJS..223...29V} Survey
of K-band selected galaxies using the VIMOS spectrograph mounted on ESO's Very Large Telescope.
The last few spectroscopic redshifts are from \citet{refId100}, who
used VIMOS, \citet{2006ApJ...644..100P} and \citet{2009ApJ...696.1195T}
who used the Hectospec multiobject spectrograph and IMACS, respectively, on the
MMT 6.5 m telescope.

For photometric redshifts, we primarily use the photometric part of the 
PAUS \citep{2021MNRAS.501.6103A} catalogue and the $K_\mathrm{s}$ selected photometric catalogue from \citep{Muzzin_2013} which contains photometry from $0.15-24\mu$m.
Then we add the Galaxy And Mass Assembly 10h region \citep[G10;][]{2015MNRAS.447.1014D}, which
contains the best photometric redshifts from PRIMUS and
\citet{2009ApJ...690.1236I}.
We also use photometric redshifts from the following works : 
the near-infrared selected
photometric catalogue created using imaging from various telescopes
covering more than 30 photometric bands \citep[hereafter COSMOS2015;][]{2016ApJS..224...24L};
the CANDELS-COSMOS survey \citep{2017ApJS..228....7N}, which is
conducted with 42 photometric bands ranging from 0.3 to 8 $\mu$m;
a volume-limited and mass 
complete sample of COSMOS2015 \citep{2017ApJ...837...16D} and
the updated edition of the Chandra COSMOS-Legacy (C-COSMOS) Survey \citet{2016ApJ...817...34M}.
In addition to the catalogues mentioned above, we used data from
SDSS-III DR12 \citep{2015ApJS..219...12A} to obtain the photometric redshifts
for our UVW1 sources.

We tabulate all these catalogues in Table \ref{tab:redshift} with the 
number of redshifts taken from each catalogue and the quality flags used
to constrain each catalogue to the best possible (most secure) redshifts.
In total, we get 4578 redshifts, of which 3658 ($\sim$ 80 percent) are spectroscopic. 

\begin{table}
  \caption{Catalogues used for spectroscopic and photometric 
  redshifts, along with the number of redshifts and quality flags 
  (QF) used for each catalogue.}
  \label{tab:redshift}
  \begin{tabular}{llr}
    \hline\hline
    \noalign{\vskip 0.5mm}
    Source Catalogue &
    Number$^a$ &
    QF \\
    \hline
    \noalign{\vskip 0.5mm}
    \multicolumn{1}{@{}l}{Spectroscopic}\\
    \citet{2018ApJS..234...21D} & 1805  & \texttt{e\char`_Z} $< 0.0002$\\
    \citet{2021MNRAS.501.6103A} & 983   & 2     \\
    \citet{2012ApJ...753..121K} & 317   & \texttt{mult}$^{a}$\\
    \citet{2018ApJ...858...77H} & 67    & 2, $\geq$ 1.5$^{b}$\\
    \citet{2019ApJS..241...10K} & 13    & $>$ 1       \\
    \citet{2011ApJ...741....8C} & 459   & 1$^{c}$   \\
    \citet{2016ApJS..223...29V} & 11    & \texttt{mult}$^{d}$  \\
    \citet{refId100}            & 1     & \texttt{none} \\
    \citet{2009ApJ...696.1195T} & 1    & \texttt{q\char`_z} $> 2$ \\
    \citet{2006ApJ...644..100P} & 1     & \texttt{none}   \\
    \hline
    \noalign{\vskip 0.5mm}
    \multicolumn{1}{@{}l}{Photometric}\\
    \citet{2021MNRAS.501.6103A}    & 559  & \texttt{none}  \\
    \citet{2015MNRAS.447.1014D}    & 35   & $\leq 2 $  \\ 
    \citet{Muzzin_2013}            & 198  & 1       \\
    \citet{2016ApJS..224...24L}    & 2   & \texttt{z68}$<0.1^{e}$      \\
    \citet{2015ApJS..219...12A}    & 122  & \texttt{e\char`_zph}$<0.1^{f}$\\
    \citet{2017ApJ...837...16D}    & 1    & 0       \\
    \citet{2016ApJ...817...34M}    & 3    & \texttt{none}   \\
    \hline
  \end{tabular}\\
  \begin{minipage}{\columnwidth}
      \textsuperscript{$a$}{Multiple constraints : \texttt{cc} $=$
      \texttt{x}.5, where \texttt{x} = 3, 4, 9, 13, 14, 18 on the first selection 
      and then, \texttt{cc} $=$ \texttt{y}.2$-$\texttt{y}.4, where \texttt{y} = 3, 
      4, 9, 13, 14, 18, 23, 24, 29 and \texttt{cc} $=$ \texttt{z}.5, where 
      \texttt{z} = 2, 12, 22.} \\
      \textsuperscript{$b$}{We use the second quality Flag for the second run.}\\
      \textsuperscript{$c$}{We choose this quality flag for the first run, and then
      on the second run, we remove this condition and 
      use secondary targets as well.}\\
      \textsuperscript{$d$}{\texttt{f\char`_z}$ =0$ and \texttt{f\char`_spec}$ =0$ 
      and \texttt{f\char`_use}$ =1$.}\\
      \textsuperscript{$e$}{\texttt{z68} represents the 68 percent confidence interval around each photometric redshift.}\\
      \textsuperscript{$f$}{\texttt{e\char`_zph}$ <0.1$ and \texttt{q\char`_mode}$ =+$ 
      and \texttt{mode}$ =1$ and \texttt{Class}$ =3$.}
  \end{minipage}
\end{table}

\begin{figure}
  \hspace*{-0.1cm}\centering
  \includegraphics [width=0.95\columnwidth]{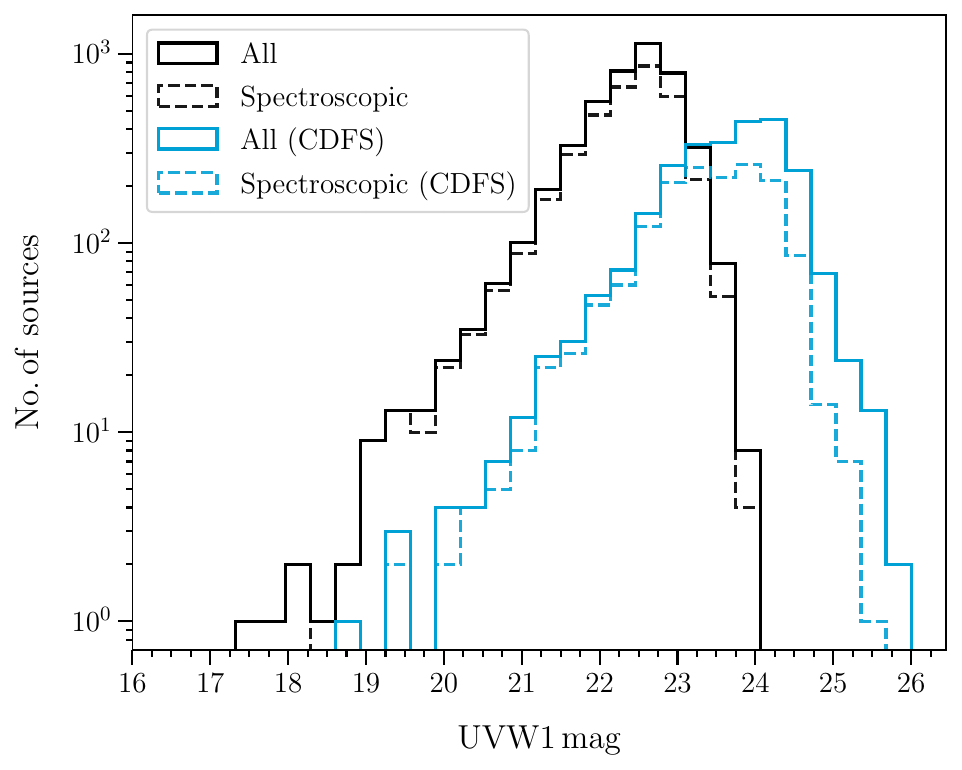}
  \caption{The distribution of the COSMOS and CDFS UVW1 sources as a function of their UVW1 magnitudes. The solid histogram shows all sources with
  redshifts and the dashed histogram represents the sources with spectroscopic redshifts.}
  \label{fig:mag_distro}
\end{figure}

\begin{figure}
  \hspace*{-0.1cm}\centering
  \includegraphics [width=0.95\columnwidth]{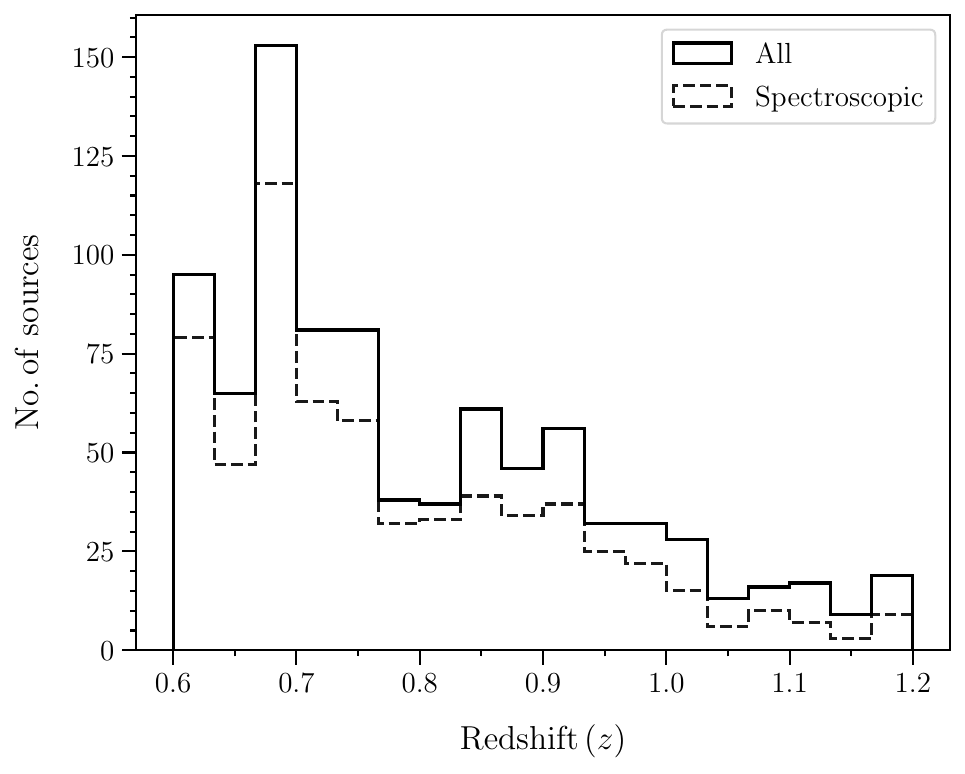}
  \caption{Redshift distribution of the COSMOS UVW1 sample. The spectroscopic redshifts are represented by the dashed line and the sum of spectroscopic and photometric redshifts is represented by the solid line.}
  \label{fig:z_distro}
\end{figure}

\subsubsection{AGNs}
\label{sec:2.4.3}

The UV contribution from the central supermassive black holes in active galaxies may dominate the emission coming from the star-forming regions,
in a UV survey. It is the latter emission that we are concerned with in this study.
Thus, the inclusion of UV-bright AGN in the sample can induce an
overestimation in the UV LF calculations. In particular, they affect the 
bright end of the UV LF. 
We use the same method as in \citetalias{2022MNRAS.511.4882S}, i.e. we
use an X-ray luminosity cut to handle the quasar contamination.
The X-ray catalogues from \citet{2016ApJ...817...34M, 2009ApJ...696.1195T} and \citet{2006ApJ...644..100P} are used to identify any X-ray sources
in the sample. Any sources cross-correlated with \citet{2016ApJ...817...34M} and having  a luminosity
greater than $10^{42}$ ergs sec$^{-1}$ in the $0.5 - 10$ KeV, 
$0.5 - 2$ KeV or $2 - 10$ KeV X-ray bands, were removed.
In addition to this, sources identified as quasars by \citet{2006ApJ...644..100P} or identified as having broad-line
features by \citet{2009ApJ...696.1195T} were also removed. 
The above criteria identify 97 sources in the UVW1 source list as 
quasars. We remove all these sources from the further analysis. It is important to remark about the possibility of some AGN making their 
way into the final catalogue, despite the luminosity cut described here, 
because the X-ray observations are not sufficiently deep to detect all AGNs with X-ray luminosities $>10^{42}$~erg~s$^{-1}$ in our redshift range.
We perform more
tests to characterise these bright UV sources in Section \ref{sec:4.6}.

After this step, we have 4481 sources in our source list. The
UVW1 magnitude distribution of this source list is plotted in Fig. \ref{fig:mag_distro}. 
Finally, a subsample of 879 galaxies has been selected with
redshifts of the highest quality within a range of $0.6 - 1.2$, 637 of which are spectroscopic. The final redshift distribution of this subsample is shown in Fig. \ref{fig:z_distro}.

\section{Corrections}
\label{sec:3}

Before the UVW1 magnitudes are converted into 1500 {\AA} absolute magnitudes and used for calculating the LFs, we need to calculate two
important corrections, that we will apply to the absolute magnitudes.

\subsection{Galactic extinction}
\label{sec:3.1}

\begin{table}
  \setlength{\tabcolsep}{10pt}
  \centering
  \caption{The first five rows of the UVW1 source catalogue are used in this work. The columns list the positions, redshifts (z) and the apparent
  UVW1 magnitudes. The full table is available in the machine-readable 
  form with the online version of the paper.}
  \label{tab:cat}
  \begin{tabular}{cccc}
    \hline\hline
    \noalign{\vskip 0.5mm}
    RA (J2000)  &
    DEC (J2000) &
    z    &
    UVW1 mag \\
    \multicolumn{1}{c}{deg} &
    \multicolumn{1}{c}{deg} &
    &
    \\
    \hline
    \noalign{\vskip 0.5mm}
    150.423 & 2.583 &  0.82   & 21.06  \\
    149.889 & 2.735 &  0.71   & 21.07  \\
    150.666 & 2.025 &  0.80   & 21.26  \\
    149.815 & 2.830 &  0.85   & 21.45  \\
    150.758 & 2.238 &  0.60   & 22.46  \\   
    \hline
  \end{tabular}\\
\end{table}

All extra-galactic surveys, especially in the blue part of the galaxy 
spectral energy distributions (SEDs) are affected by extinction from Galactic foreground dust.
We estimate the extinction correction for the COSMOS
source list using the Extinction Calculator\footnote{\url{https://ned.ipac.caltech.edu/extinction_calculator}} tool provided by the NASA/IPAC Extragalactic Database (NED). This tool uses the \citet{2011ApJ...737..103S}
re-calibration of the \citet{1998ApJ...500..525S} dust map.
In particular they used the extinction law parameterised as $R_\mathrm{V} = A_\mathrm{V} / E \left(B - V\right)$ and calibrated it assuming an $R_\mathrm{V} = 3.1$ \citep{1989ApJ...345..245C}. We assume an effective wavelength of 2934.5 {\AA} and find a value of 0.097 magnitudes for Galactic extinction in the UVW1 band towards the centre of the COSMOS UVW1 image.
We note here that the COSMOS UVW1 image has a large area ($\sim 1.5$ sq. degrees) and shows
a variation in the dust extinction. To estimate this variation, we use the same  NED tool and calculate the attenuation in the directions of the centres of all the 25 sub-fields that constitute the COSMOS UVW1 image (see Fig. \ref{fig:uvw1_im}) and tabulate them in Table \ref{tab:e_b_v} of Appendix~\ref{sec:dust_ext}. The mean of these values comes out to be 0.099 with a standard deviation of 0.006.

We assume the correction for the whole field to be 0.097 mag and add it to the calculation of absolute magnitude for each source.
The sample is made available as a supplementary table in the online version of the paper. The first five rows of the table are shown in Table \ref{tab:cat}.

\subsection{K-correction}
\label{sec:3.2}

\begin{figure}
  \hspace*{-0.1cm}\includegraphics[width=0.95\columnwidth]{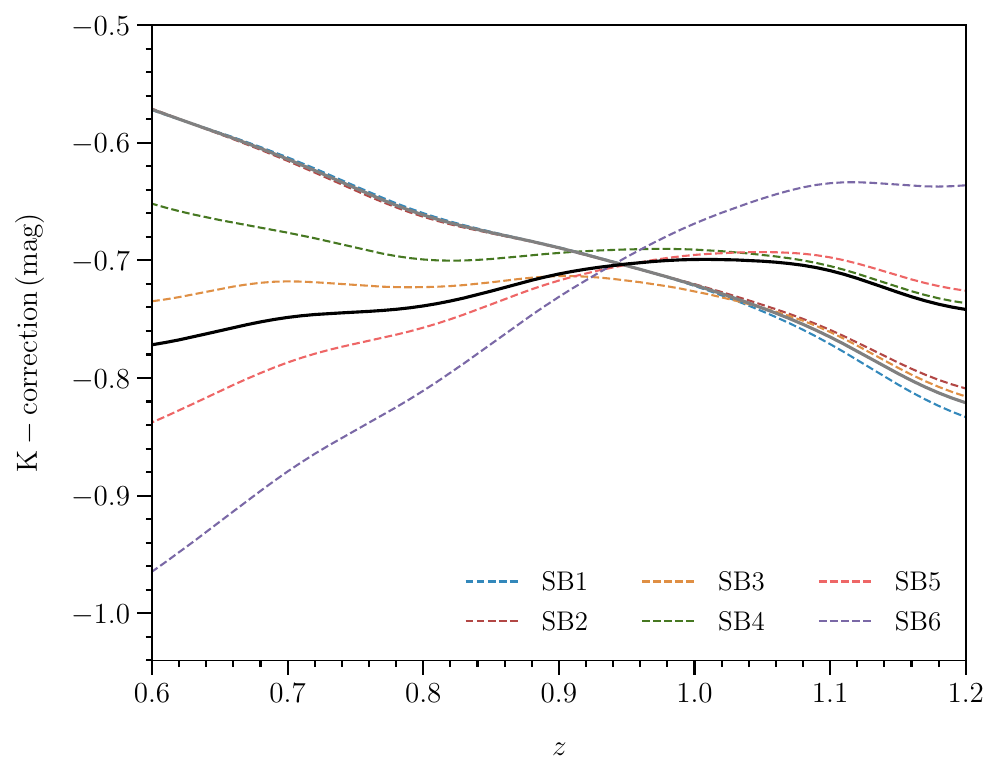}
  \caption{K-corrections as a function of their redshifts for the spectral templates of starburst galaxies from \citet{1996ApJ...467...38K} shown by dashed lines. These templates (SB1-SB6) are extended blue-ward of {\AA} using the spectrum of Mrk 66 from \citet{1998ApJ...495..698G}. The grey solid line shows the weighted average of the low extinction templates SB1-SB2, whereas the black solid line represents the same for starburst templates SB3-SB6 (also see Section \ref{sec:6.4}).}
  \label{fig:kcorr}
\end{figure}

The observed SED of a galaxy appears different from the rest-frame SED 
as it gets red-shifted due to the expansion of the Universe.
So, depending upon the redshift of a source a single waveband may look at different 
parts of the galaxy SEDs. This will affect the calculation of
the absolute magnitude of a source in that given waveband and cause
erroneous results. 
To avoid this effect we have to add a compensating
term to the absolute
magnitudes of the galaxies in our sample. This compensating term is 
called the K-correction \citep{2002astro.ph.10394H}, 
defined as the ratio of fluxes in the rest-frame waveband for which we are calculating an absolute magnitude and the observed band in which we have measured an apparent magnitude.
We summarise here the methodology used to calculate
the K-correction which is described in \citet{2021MNRAS.506..473P} in detail.
Due to a shallow survey in the COSMOS field, we expect the sources to be bright starburst
galaxies. So, to quantify their K-corrections the appropriate models to use
are the empirical starburst templates from \citet{1996ApJ...467...38K}.
These templates do not extend to wavelengths below 1,200 A, but for $z>1$ galaxies the UVW1 filter is sufficiently wide that it includes some emission at rest-frame wavelengths below 1,200 {\AA}. To calculate the K-corrections at these redshifts, we extend the 
starburst templates blue-ward of 1200 {\AA} by using the spectrum of Mrk 66 from \citet{1998ApJ...495..698G}.
As our UV selection will pick up dust-free bright UV galaxies, we expect that the 
K-corrections corresponding to the lowest extinction template SB1 should be appropriate for 
our study. However, we do test this assumption by dividing our sample into two sub-samples
based on their $B-I$ colour. We discuss these tests in detail in Section \ref{sec:6.4}.
The calculated corrections are plotted in Fig. \ref{fig:kcorr}.
As in \citetalias{2022MNRAS.511.4882S}, the K-corrections $K\left(z\right)$ are added to
the expression calculating the 1500 {\AA} absolute magnitudes from the apparent UVW1 magnitudes 
(Equation \ref{eq:absmag}). We use the \textit{GALEX} FUV passband to define the 1500 {\AA} absolute magnitudes.
\begin{equation}
\label{eq:absmag}
M_{1500}\left(z\right) = m-5\log{\left(\frac{d_L\left(z\right)}
    {\mathrm{Mpc}}\right)}-25-K\left(z\right)-X_{\mathrm{UVW1}},
\end{equation}
where $X_{\mathrm{UVW1}}$ is the extinction correction from 
Section \ref{sec:3.1} and $d_L\left(z\right)$ is the luminosity distance.

\section{Luminosity function and Luminosity density}
\label{sec:4}
Once all corrections have been applied to the sample, we divide the sample into two redshift
bins covering $0.6 < z < 0.8$ and $0.8 < z < 1.2$. After this we calculate the binned LF, estimate the Schechter function parameters and in the end, determine the UV luminosity density.
This Section outlines the formalism for all these calculations. We also outline
the method we employ to construct the spectral energy distributions (SEDs)
for the brightest sources.

\subsection{Binned luminosity function estimate}
\label{sec:4.1}

There are many popular methods to estimate the binned luminosity function e.g. the 1/$V_{\mathrm{max}}$ method \citep[][]{1968ApJ...151..393S}, the 1/$V_{\mathrm{a}}$ method
\citep[][]{1996MNRAS.280..235E} etc. \citep[see][for a review]{2011A&ARv..19...41J}.
These methods have a significant systematic error for objects close to the flux limit
\citep{2000MNRAS.311..433P}, so we use an improved estimator from \citet{2000MNRAS.311..433P}.

In volume-magnitude space, we define the binned luminosity
function as the number of galaxies $N_{\mathrm{bin}}$ inside the
bin bound by the redshift interval $z_\mathrm{min}<z<z_\mathrm{max}$ 
and magnitude interval $M_\mathrm{min}<M<M_\mathrm{max}$, divided
by the effective survey volume enclosed by that bin
\citep{2000MNRAS.311..433P},
\begin{equation}
  \phi\left(M,z\right) \equiv \phi = \frac{N_{\mathrm{bin}}}{V_{\mathrm{bin}}}.
\end{equation}
The effective survey volume inside each bin $V_\mathrm{bin}$ is a 
4-volume in the volume-magnitude space given by,
\begin{equation}
  V_{\mathrm{bin}} = 
  \int_{M_\mathrm{min}}^{M_\mathrm{max}}\,
  \int_{z_{\mathrm{min}}}^{z_{\mathrm{max}}}
  \frac{\mathrm{d}V\left(z\right)}{\mathrm{d}z}\, 
  \mathrm{d}z\, \mathrm{d}M, 
  \label{eqn:iv}
\end{equation}
where $z_\mathrm{min}$ and $z_\mathrm{max}$, the lower and upper 
extremes of the redshift interval are calculated as
\begin{equation}
  z_{\mathrm{min}} = \mathrm{min}\left(z_{l},\, z^\prime\left(M,\,
  m_{l}\right)\right),
\end{equation}
and
\begin{equation}
  z_{\mathrm{max}} = \mathrm{max}\left(z_{u},\, z^\prime\left(M,\,
  m_{u}\right)\right),
\end{equation}
where $m_{l}$ and $m_{u}$ are the lower and upper magnitude limits of
the survey, $z_{l}$ and $z_{u}$ are the lower and upper limits of the redshift bin, $z^\prime\left(M,\, m_{l}\right)$ and $z^\prime\left(M,\, m_{u}\right)$ are the minimum and maximum redshifts such that which the object can be found and still be within the magnitude limits of
the survey.
The term $\mathrm{d}V\left(z\right)/\mathrm{d}z$ in the integrand of Equation \ref{eqn:iv} is obtained as a product of the effective area and the differential
comoving volume elements,
\begin{equation}
  \frac{\mathrm{d}V\left(z,M\right)}{\mathrm{d}z} =
  \left(\frac{\pi}{180}\right)^2\,
  \int_{\Omega}\,  \mathcal{A}_{\mathrm{eff}}\left(M\right) \,  
  \frac{\mathrm{d}V\left(z\right)}{\mathrm{d}z\, \mathrm{d}\Omega}\,
  \mathrm{d}\Omega\, 
\end{equation}
where $\mathcal{A}_{\mathrm{eff}} = \mathcal{A} \cdot C\left(m\right)$ is the effective 
area, obtained by 
multiplying the sky-area ($\mathcal{A}$) by the completeness function $C\left(m\right)$. We tabulate effective areas along with
completeness as a function of the UVW1 magnitudes in Table \ref{tab:comp}. 
\begin{equation}
  \frac{\mathrm{d}V\left(z\right)}{\mathrm{d}z\, \mathrm{d}\Omega\,} = 
  \frac{c\,H_{0}^{-1}\,d_{L}^{2}}
  {(1+z)^2\,[\Omega_{\lambda} + \Omega_{m} (1+z)^3]^{1/2}}\,
\end{equation}
is the differential comoving volume element \citep[][]{1999astro.ph..5116H}.

\begin{table}
  \centering
  \caption{The COSMOS UVW1 survey has different levels of completeness at different magnitudes, hence the effective sky area changes with UVW1 magnitude. 
  Using the completeness simulations we associate UVW1 magnitudes (first column) with
  completeness fractions (second column). For each magnitude, a completeness fraction is estimated as explained in section \ref{sec:2.3}.
  The effective area ($\mathcal{A}_{\mathrm{eff}}$) is the product of the completeness fraction and the geometric sky area of the survey
  and is given in the third column.
}
  \label{tab:comp}
  \begin{tabular}{ccccccr}
    \hline\hline
    \noalign{\vskip 0.5mm}
    \multicolumn{1}{c}{UVW1 magnitude} &
    \multicolumn{1}{c}{Completeness} &
    \multicolumn{1}{c}{Effective Area} \\
    \multicolumn{1}{c}{(AB mag)} &
    \multicolumn{1}{c}{(percent)} &
    \multicolumn{1}{c}{(deg$\mathrm{^2}$)} \\
    \hline
    \noalign{\vskip 0.5mm}
    19.93 & 98.83 & 1.5088 \\
    21.66 & 95.30 & 1.4550 \\
    22.16 & 88.04 & 1.3441 \\
    22.32 & 80.36 & 1.2269 \\
    22.41 & 73.24 & 1.1182 \\
    22.56 & 61.17 & 0.9339 \\
    22.66 & 52.44 & 0.8007 \\
    22.76 & 39.90 & 0.6094 \\
    22.86 & 30.65 & 0.4679 \\
    22.93 & 25.26 & 0.3855 \\
    23.02 & 17.76 & 0.2711 \\
    \hline
  \end{tabular}\\
\end{table}

From Poisson's statistics \citep{1986ApJ...303..336G}, we calculate
the uncertainty for $N$ objects and hence the statistical uncertainty in
the LF for each bin.
The resulting luminosity function $\phi$ has units of
$\mathrm{Mpc}^{-3}\mathrm{mag}^{-1}$.

\subsection{Schechter function parameters}
\label{sec:4.2}

We analyse the galaxy distribution in the redshift
-magnitude space by comparing it to a galaxy LF model using maximum
likelihood. 
Similar techniques were first applied to galaxy luminosity functions by \citet{1979ApJ...232..352S}.
We use the Schechter function \citep{1976ApJ...203..297S} 
to model the galaxy LF in each redshift bin. It is parametrised as,
\begin{multline}
    \phi(M) = 
    0.4\, \ln{10}\, \phi^{*} \\
    \times 10^{-0.4(M-M^{*})(1+\alpha)}\,
    \mathrm{exp}\left({-10^{-0.4(M-M^{*})}}\right)
    \label{eqn:phi}
\end{multline}
the product of an exponential function and a power law, which dictate the shape at bright and faint magnitudes respectively.
The shape transitions from the power law of slope $\alpha$, to the
exponential form at the knee of the LF, described by $M^{*}$.
The LF is normalised by the characteristic number density
$\phi^{*}$.\\ 
We convolve the Schechter function model with the error distributions of magnitudes ($P\left(M^\prime|M\right)$; where $M$ and $M^\prime$ are the true and observed magnitudes; see Fig. \ref{fig:hist}) obtained from the completeness
simulations in Section \ref{sec:2.3}. These histograms are normalised
by the number of sources for each magnitude that are inserted into
the image (see \citet{2021MNRAS.506..473P} and \citet{2022MNRAS.511.4882S} for details on the propagation of measurement uncertainties in the estimation of the LF parameter), so completeness at the corresponding magnitude is accounted for.

The probability distribution such that a galaxy $i$ of magnitude $M$ is
observed with a magnitude $M^{'}_i$ at a redshift $z_{i}$, can be expressed as,
\begin{multline}
p\left(M^\prime_{i},z_{i}\right) = \\
\frac{\displaystyle
\int_{M_{\mathrm{min}}}^{M_{\mathrm{max}}}
\phi\left(M, z_i\right) P\left(M^\prime_i|M\right)\mathrm{d}M}
{\displaystyle 
\int_{M^\prime_{\mathrm{min}}}^{M^\prime_{\mathrm{max}}}
\int_{z_{\mathrm{min}}}^{z_{\mathrm{max}}}
\int_{M_{\mathrm{min}}}^{M_{\mathrm{max}}}
\phi\left(M, z\right) P\left(M^\prime|M\right)\mathrm{d}M\,
\frac{\mathrm{d}V \left(z\right)}
{\mathrm{d}z}\,\mathrm{d}z\, \mathrm{d}M^\prime}
\label{eq:p}
\end{multline}
where the $M_{min}$ and $M_{max}$ define the range of absolute magnitudes that could plausibly give rise to the measured absolute magnitude $M_{i}’$, given the magnitude error distribution. Note that $M_{min}$ and $M_{max}$ in Equation \ref{eq:p} are unrelated to the delimiters of the absolute magnitude bins employed in the binned luminosity function estimation despite our use of the same symbols for that purpose in Section \ref{sec:4.1}. 

Given the LF parameters $\theta = \left(\phi^{*},M^{*},\alpha\right)$, 
the probability distribution can be written as,
\begin{equation}
    p\left(M^\prime_{i},z_{i}\,|\,\theta\right)\propto
    \frac{p\left(\theta\,|\,M^\prime_{i},z_{i}\right)}
    {p\left(\theta \right)}.
    \label{eqn:p2}
\end{equation}

Then the likelihood function for $N_{G}$ galaxies will be,
\begin{equation}
  \mathcal{L}= \prod_{i}^{N_{G}}p\left(M^\prime_i, z_i\,|\,
  \theta\right)
\end{equation}
or in a more convenient form as
\begin{equation}
  S = -2\ln\mathcal{L} = 
  -2 \sum_{i=1}^{N_\mathrm{G}}\ln p\left(M^\prime_i, z_i\right),
  \label{eqn:L}
\end{equation}
and it can be minimised to maximise $\mathcal{L}$. 
The normalisation $\phi^{*}$ gets cancelled out in \ref{eq:p}, so it cannot be determined using this formalism alone. It is determined by additionally asserting the condition that
\begin{equation}
    \iint
    \phi\left(M^\prime_i, z_i\right)
    \frac{\mathrm{d}V \left(z\right)}
    {\mathrm{d}z}\,\mathrm{d}z\, \mathrm{d}M^\prime
    = N_{G}
    \label{eq:p3}
\end{equation} 
i.e. equating the predicted and observed number of sources.
We obtain the posterior probability distributions for the parameters of the Schechter function $\left(\phi^{*},\, M^{*},\, \alpha\right)$ using
the Markov Chain Monte Carlo (MCMC) method, assuming a uniform non-informative prior $p\left(\theta \right)$.
To implement MCMC, we use \textsc{python} module \textsc{emcee} \citep{2013PASP..125..306F}.

\subsection{UV Luminosity density}
\label{sec:4.3}

Once the Schechter function parameters are determined for our survey,
the luminosity density can be derived by integrating the product of
luminosity with the luminosity function,
\begin{equation}
  j =  \int_{0}^{\infty} L\, \phi\left(L\right)\,
  \mathrm{d}L,
  \label{eqn:j}
\end{equation}
using Equation \ref{eqn:phi} with 
$L/L^{*} = \mathrm{exp}[-0.4\ln{10}\cdot(M - M^*)]$, we can write
\begin{equation}
  j =  \int_{0}^{\infty} L\, \phi^{*}
  \left(L/L^{*}\right)^{\alpha}\,
  \mathrm{exp}\left(-L/L^{*}\right)\,\mathrm{d}\left(L/L^{*}\right).
  \label{eqn:j2}
\end{equation}
This gives us a more robust quantity than normalisation ($\phi^*$) of 
the LF to compare our results with past 
works and avoids the degeneracy in $\phi^{*}$ and $M^*$.
We integrate from $M_{1500} = -10 \, (L_{1500} = \num{4.3e24})$ to 
$M_{1500} = -24 \, (L_{1500} = \num{1.7e30})$ to get the luminosity density.
The errors on the normalisation due to cosmic variance (Section \ref{sec:4.5}) are included in addition to the statistical errors
in our calculation of the luminosity density.

\subsection{Cosmic Variance}
\label{sec:4.5}

The LF estimates are prone to errors due to the large-scale matter distribution
in the Universe. Due to the variation in the matter
density, the number counts of the galaxies fluctuate from one part of the 
Universe to another. This effect is most severe for
surveys with small sky areas  \citep{2004ApJ...600L.171S,2010ApJ...710..903M}.
In this Section, we calculate the effects of this so-called cosmic
variance on our estimates.
We use two independent methods, proposed by \citet{2008ApJ...676..767T}
and \citet{2010ApJ...710..903M} to obtain cosmic variance-induced errors
on the characteristic number density (or the normalisation) $\phi^*$ 
of the Schechter function form of the LF.
The first one introduced by \citet{2008ApJ...676..767T}, calculates cosmic variance using an approach in which the N-body simulations are used to produce mock surveys which in turn
are used to calculate the average bias of the sample.
The other method suggested by \citet{2010ApJ...710..903M}, estimates the cosmic variance using the N-body simulated mocks from the sky area of the survey, the mean and size of the redshift bin, and the stellar masses of the galaxies probed.

In the web tool provided by \citet{2008ApJ...676..767T}, we assume values for $\sigma_{8}$ and an average halo occupation fraction
of 0.8 and 0.5, respectively, along with the bias formalism of
\citet{1999MNRAS.308..119S}. It gives us $1 \sigma$ fractional errors
of 0.10 and 0.08 on normalisation.

As mentioned earlier, the method from \citet{2010ApJ...710..903M}
needs the stellar masses. So, we match our final UVW1 source-list
to COSMOS/UltraVISTA $K_{s}-$selected catalogue \citep{Muzzin_2013} with 
a matching 
radius 1.5 arcsec to get the stellar masses of our galaxies.
The matching provides 495 stellar masses (96.5 percent of our sources), 
which average to $\num{1.51e10}{M_\odot}$ in the redshift bin $0.6-0.8$. 
For these stellar masses, we get
a relative $1 \sigma$ error of 0.11 from the \citet{2010ApJ...710..903M} code.
Following the same procedure for the redshift bin $0.8-1.2$, we get
349 (95.4 percent) counterparts with stellar masses for our sources, 
giving an average stellar mass $\num{1.87e10}{M_\odot}$ and a relative $1 \sigma$ error on normalisation of 0.07, due to cosmic variance.
The estimated values of relative errors due to cosmic variance were calculated using tools from both
\citet{2008ApJ...676..767T} and \citet{2010ApJ...710..903M} are tabulated 
in Table \ref{tab:cv}.

Cosmic variance is known to be correlated with the bias present in the galaxy population under consideration \citep[][]{2010ApJ...710..903M}. It has been noted that galaxies with lower UV luminosities may be slightly less biased than their brighter counterparts \citep{2009A&A...498..725H,2018A&A...614A.129V}, thus leading to a reduction in the Cosmic Variance at lower luminosity levels.

Given the wide range of luminosities within our sample,
we conducted a preliminary analysis within each redshift bin to investigate any correlation between cosmic variance estimates and source brightness. 
Within each redshift bin, we categorized sources into two groups based on their $M_{\mathrm{UVW1}}$ values: \textit{bright} and \textit{faint}. For the redshift bin $0.6-0.8$, sources with $M_{\mathrm{UVW1}}$ values below $22.75$ were classified as bright, while those with values above this threshold were classified as faint. In the other redshift bin ($0.8-1.2$), the threshold of $M_{\mathrm{UVW1}}$ was set at $22.85$. The threshold values were carefully chosen to ensure that the number of objects in both bright and faint groups is roughly similar.

We do not observe any significant change in the average value of cosmic variance when comparing the entire sample to the faint subsamples in both redshift bins. However, in the case of the bright sample, we observed a 10 percent increase in the average cosmic variance relative to the entire sample in both redshift bins.
The statistical error bars in normalisation (see Table \ref{tab:par_f}) are considerably larger than those associated with cosmic variance. Even after accounting for the aforementioned 10 percent variation in cosmic variance errors, statistical uncertainties remain the dominant factor in our error budget.

\subsection{Spectral energy distributions}
\label{sec:4.6}

We fit spectral
energy distributions (SEDs) to the galaxies in the brightest magnitude bins in both redshift ranges to examine their nature.
We obtain the rest-frame photometry for our sources in different filters from UV 
to mid-IR, by matching the positions to the COSMOS2015
catalogue.
For sources that do not have photometry in the Far-IR bands, we used the deblended
photometry from \citet{2018ApJ...864...56J} and the HerMES catalogue
\citep[Herschel Multi-tiered Extragalactic Survey;][]{2012MNRAS.424.1614O}.

An SED model comprising two components is fitted to the photometry. The first component
is the stellar emissions from the
galaxies and the other component is from the dust emissions coming 
from the star-forming regions. 
The stellar emission templates are created using stellar population synthesis models
of \citet{2003MNRAS.344.1000B},
assuming the \citet{2003PASP..115..763C} initial mass function for solar metallicity. 
We chose
models of the single stellar population (SSP) with a varying age from 0.01 to 13.18
Gyr, in 30 steps of size log age $\sim$ 0.1.
These templates are reddened by assuming the  
\citet{2000ApJ...533..682C} dust extinction model, with the $E(B-V)$ values 
ranging from 0.0 to 4.0 in steps of 0.14.
In total, we have 900 stellar emission models.
We have chosen the SSP population for simplicity. The sources we have at hand are very luminous in the UV, and have blue colours, and so their stellar emission is likely to be dominated by the light from young stars.
We complement these with the mid- to far-infrared models created by
\citet{2007ApJ...657..810D}.
The cold dust component of our SED constitutes a linear combination of 
models with constant and variable radiation fields, along with varying 
amounts of the PAH fraction. Since we are dealing here with bright/large
galaxies, so the SMC and LMC models from \citet{2007ApJ...657..810D} are not 
included in our library.
From the SED fits, we calculate their total far-infrared luminosities by integrating
the total dust model from 8 to 1000 $\micron$. Bolometric luminosity is
obtained by integrating the total SED fit within $ 0.01-1000\micron$.

\section{Results and Discussion}
\label{sec:6}

The galaxy rest-frame UV LF in the redshift range $0.6 < z < 1.2$ is
derived using the method developed by \cite{2000MNRAS.311..433P}. We produce our results by dividing the sample into two redshift bins $0.6 < z < 0.8$ and $0.8 < z < 1.2$. 
We calculate the UV LF of galaxies using UVW1 data from the wide area COSMOS 
field. This work complements \citetalias{2022MNRAS.511.4882S}, in which LFs were calculated from deep UVW1 imaging in the CDFS, in that the larger sky area of COSMOS provides access to a larger sample of the most luminous galaxies, and so we can construct LFs to brighter absolute magnitudes.

We present the colour properties (Section \ref{sec:6.1}) of the UVW1 COSMOS sample before
the LF results. The colour properties are studied to assess the impact of our choice of the \citet{1996ApJ...467...38K} starburst templates, used for K-correcting the UVW1 magnitudes, on the
LF calculations.
The binned LF and the Schechter function fit to the UVW1 sources are presented in Section \ref{sec:6.2}.
We discuss the shape of the LF in the same Section. In Fig. \ref{fig:lf68comp1}, we show the comparison of the results obtained by this study and from \citetalias{2022MNRAS.511.4882S}, for
redshift bins centred at $z=0.7$ and $z=1.0$ respectively. The top panels compare the binned and
model LF whereas the bottom panels compare the parameter space, for both redshift bins.
The Schechter function parameters are compared
to estimates obtained by previous work in Section \ref{sec:6.3} and plot the comparison in Fig. \ref{fig:comp_par}. Our estimates for the luminosity density are plotted with values
from the literature in Fig. \ref{fig:lum_dens}, and tabulated in Table
\ref{tab:lumden}.
In the end, in Section \ref{sec:6.4}, we discuss the evolution of the LF and the luminosity density.

\subsection{Colour properties}
\label{sec:6.1}

\begin{figure}
  \hspace*{-0.1cm}\includegraphics[width=0.95\columnwidth]{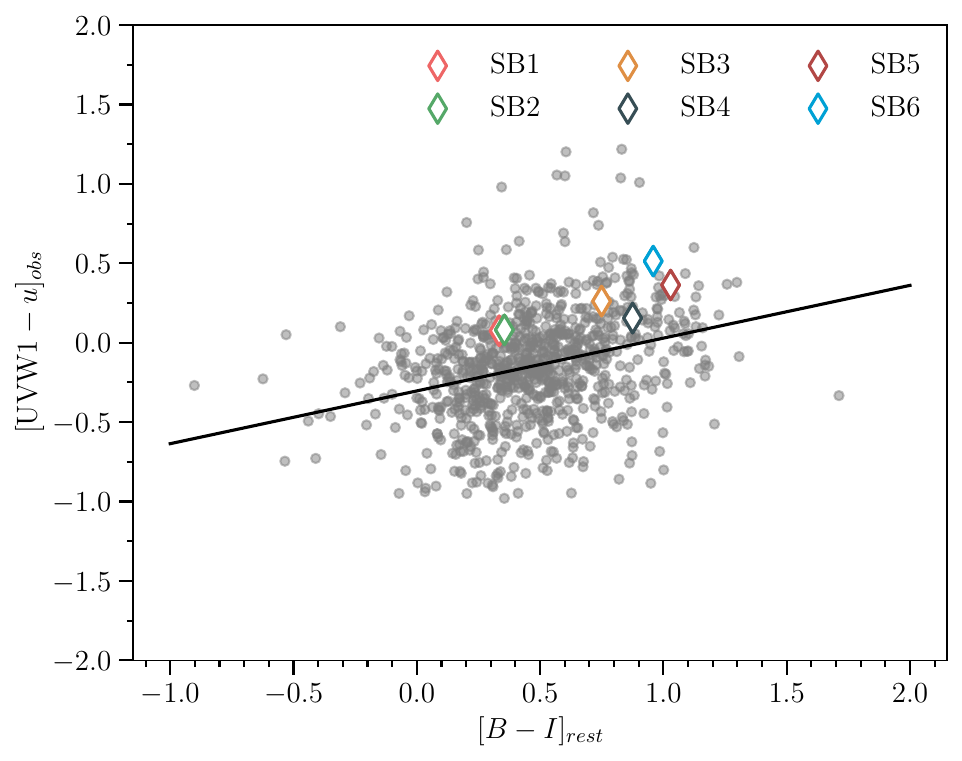}
  \caption{The observed $\mathrm{UVW1}-u$ vs rest-frame $B-I$ colours. The grey circles represent our UV-selected galaxies and the black solid line shows the best-fit 
  line to the data. The positions of the starburst templates from \citet{1996ApJ...467...38K} at a redshift of 0.9 are plotted as coloured diamond symbols.}
  \label{fig:uwv1_u_vs_b_i}
\end{figure}

\begin{figure}
  \hspace*{-0.1cm}\includegraphics[width=0.95\columnwidth]{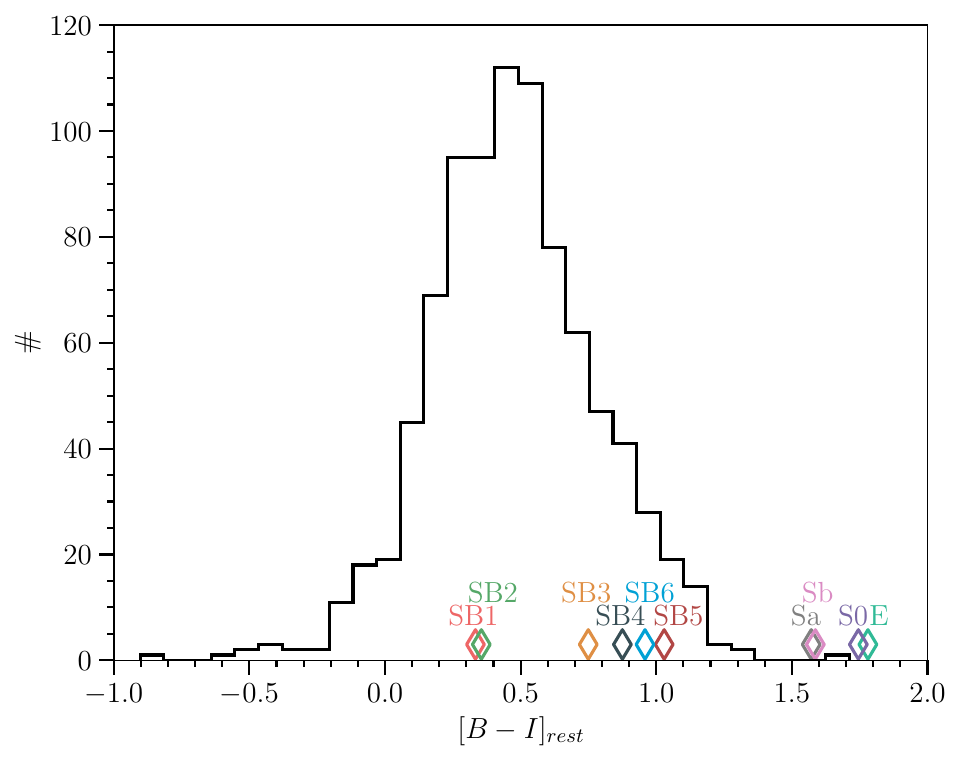}
  \caption{The distribution of rest-frame $B-I$ colours of our UVW1 sample. It is shown as the black histogram. The coloured diamonds show the $B-I$ colours calculated for different galaxy spectral types taken from \citet{1996ApJ...467...38K}.}
  \label{fig:b_i_hist}
\end{figure}

Dust within galaxies plays a significant role in shaping the UV spectral slope ($\beta$). This is because dust grains both absorb and scatter UV photons, leading to a reddening of the UV continuum. Consequently, galaxies with substantial dust content tend to exhibit a redder UV slope.
To classify our sources according to their relative dust content, we employ observed frame $\mathrm{UVW1}-u$ colours as a proxy for the UV spectral slope.
In Fig. \ref{fig:uwv1_u_vs_b_i}, we show the relationship between the observed $\mathrm{UVW1}-u$
colour and the rest-frame $B-I$ colour for our sources. The correlation suggests that the rest-frame $B-I$ colour can be utilized as an indicator of stellar reddening in the rest-frame.
It is evident from this figure that the starburst templates from \citet{1996ApJ...467...38K} follow a similar trend as the UVW1 sources. However, there is a slight offset in the $\mathrm{UVW1}-u$ colour, approximately 0.25 when compared to the best-fitting line for the UVW1 sources.
Because most of the sources are close to the magnitude limit
The majority of this offset can be attributed to bias in the UVW1 magnitudes from source detection. At the faintest limit of our survey, which is 23.02 (or 23.11 without accounting for Galactic extinction correction), as indicated by a vertical line in Fig. \ref{fig:hist2}, the bias in UVW1 magnitude is roughly 0.22.
The remaining offset suggests that our UV selection typically identifies sources with bluer UV slopes compared to the \citet{1996ApJ...467...38K} templates, for a given a specific rest-frame $B-I$ color.

Following the approach of \citet{2005ApJ...619L..43A}, we use rest-frame $B-I$ colours to differentiate between various spectral types of galaxies in our sample. 
In Fig. \ref{fig:b_i_hist}, we plot the distribution of rest-frame $B-I$ colours for our sources. 
To examine whether there is a contribution from galaxies with spectra typical of spiral galaxies of the latter type, we include the positions of these galaxies in the distribution.
As evident from Fig. \ref{fig:b_i_hist}, our sample is dominated by low-extinction galaxies and is almost completely composed of sources corresponding to the starburst templates in terms of optical colours. This is the reason why our K-corrections calculations are limited to starburst galaxies.

We divide our sample into two subgroups based on their rest-frame $B-I$ colours. The first subgroup (low-reddening) consists of sources with $[B-I]_{\text{rest}} < 0.56$, while the second subgroup (moderate-reddening) includes those with $[B-I]_{\text{rest}} \geq 0.56$ \citep{2005ApJ...619L..43A}.
The low-reddening subgroup corresponds to the low-extinction starburst templates SB1 and SB2 \citep[$E(B-V)<0.21$;][]{1996ApJ...467...38K}, whereas the moderate-reddening subgroup is populated by starburst galaxies SB3-SB6 (see Fig. \ref{fig:b_i_hist}).

For these subgroups, we calculate the binned LF using different K-corrections. 
The weighted mean of the K-corrections for templates SB3 to SB6 is applied to the moderate-reddening subsample, with weights based on the number of sources in the 0.1-magnitude interval centred around the rest-frame $B-I$ value for each starburst template. This is represented as the black curve in Fig. \ref{fig:kcorr}.
Similarly, for the low-reddening subsample, we use the weighted mean of the K-corrections for the SB1 and SB2 templates, represented as a solid grey curve in Fig. \ref{fig:kcorr}.

\begin{figure}
  \includegraphics[width=0.90\columnwidth]{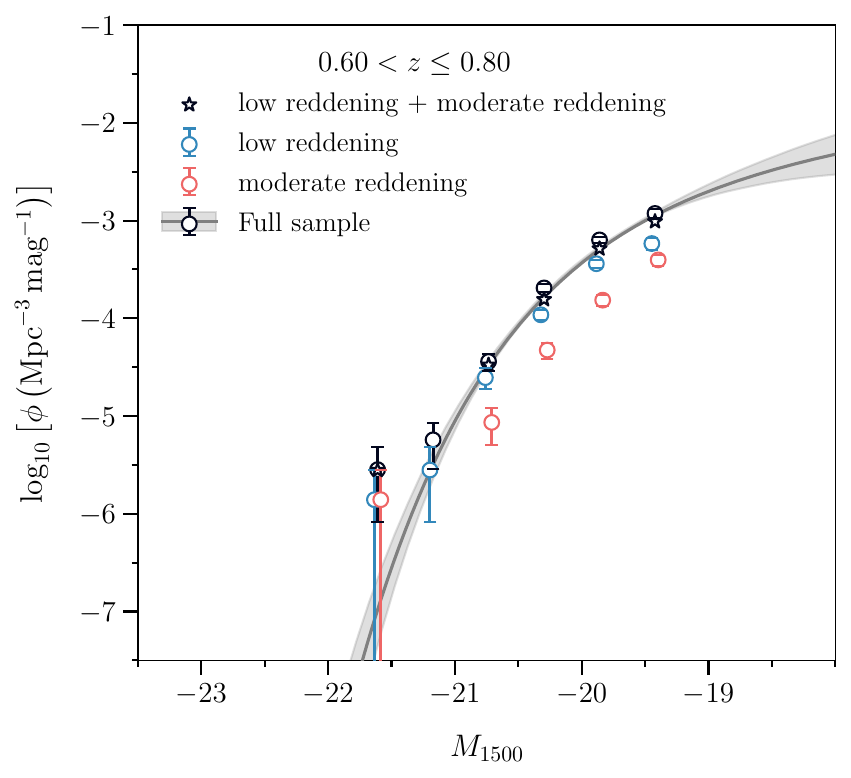}
  \includegraphics[width=0.90\columnwidth]{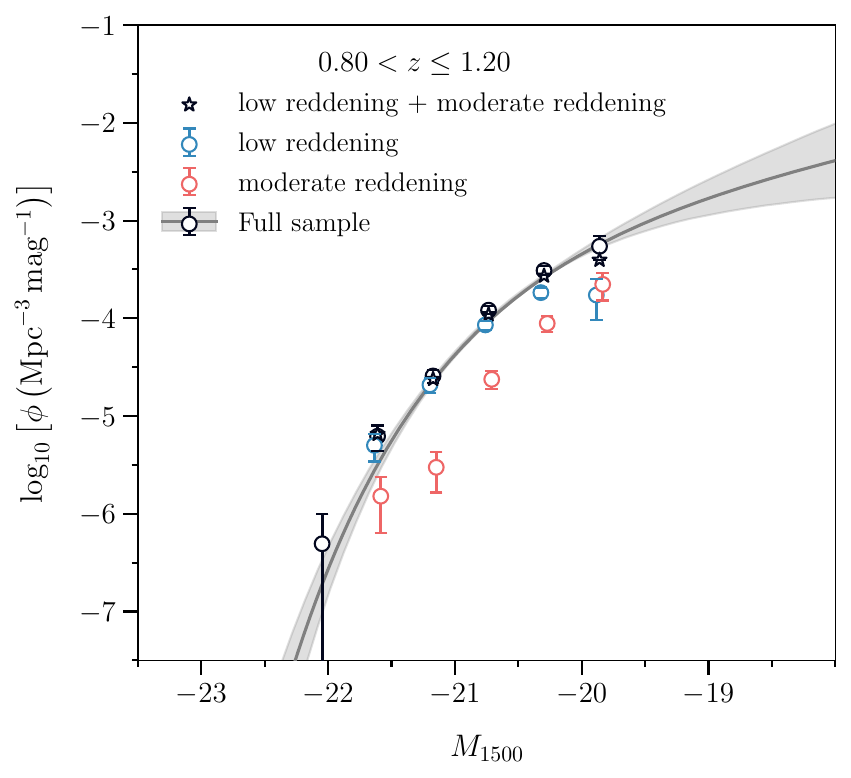}
  \caption{Here we show the comparison of the binned UV LF for the colour-divided sub-samples and our overall sample. The \textit{top} and bottom panels represent the two redshift bins $0.6-0.8$ and $0.8-1.2$.
  The black circles represent the binned UV LF from the overall sample. The blue and red coloured circles represent the binned UV LF from our low-reddening (low extinction starburst) and Sub-sample2 (high-extinction starburst). The black stars show the sum of the binned UV LF from both sub-samples. For comparison, we also plot the Schechter function fit to the overall sample.}
  \label{fig:lf_kcorr_test}
\end{figure}

\begin{figure}
  \includegraphics[width=0.95\columnwidth]{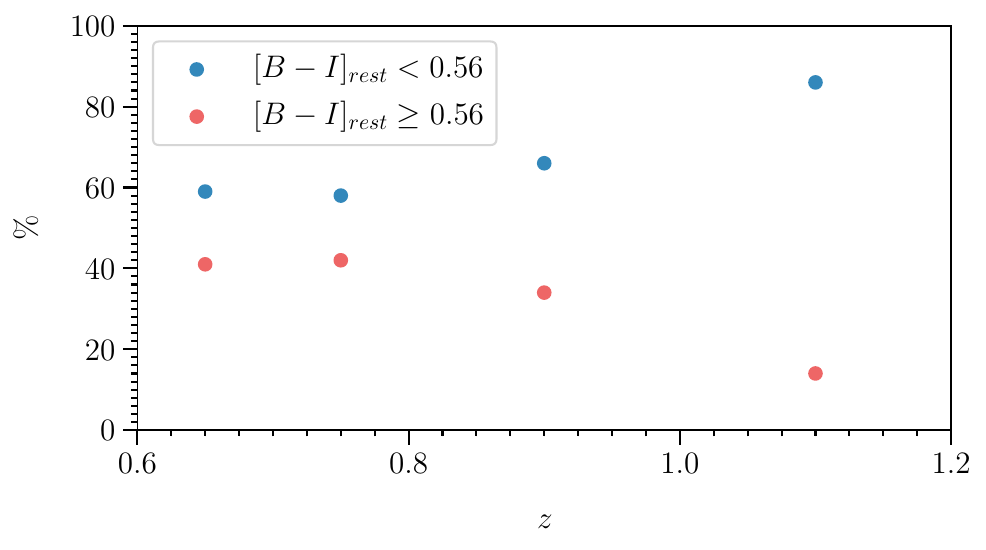}
  \caption{The fraction of galaxies from both sub-samples as a function of redshift. Blue and red coloured circles represent the fractions of low extinction (Sub-sample1) and high extinction (Sub-sample1) starbursts respectively.}
  \label{fig:gal_type_contr}
\end{figure}

The results are presented in Fig. \ref{fig:lf_kcorr_test}, where the binned LFs for these subsamples are represented by blue and red open circles, respectively.
The sum of the binned LFs from both subgroups is denoted by black stars, and they closely align with the binned LF values obtained from the entire sample (black circles). 
This suggests that the majority of galaxies in our COSMOS UVW1 sample are starbursts with minimal dust content. The low-extinction starburst template (SB1) from the \citet{1996ApJ...467...38K} library can be used effectively to K-correct our sample.
In both of the redshift bins, we can observe an interesting trend in the LF of our subsamples categorized by varying levels of reddening. 
Specifically, at bright magnitudes the LF for galaxies with moderate reddening is significantly lower than the LF of the low-reddening subsample. However, as we move towards fainter magnitude bins at both redshifts (0.7 and 1.0), the LFs of these two subsamples gradually converge.
This observation is interesting because it suggests that our LFs are dominated by galaxies with low levels of reddening, except at the lowest luminosities probed in this study. It is at this point that galaxies with significant reddening begin to contribute significantly to shape the LF.
This trend aligns with a scenario in which galaxies affected by reddening exhibit a LF shifted towards lower luminosities compared to their unreddened counterparts, a consequence of the reddening process.

In Fig. \ref{fig:gal_type_contr}, we present the fraction of galaxies from each subgroup within four redshift bins: $0.6-0.7$, $0.7-0.8$, $0.8-1.0$, and $1.0-1.2$. 
It is evident that the fraction of galaxies from low reddening (low-extinction starbursts) subsample increases with redshift, rising from 59 percent in the redshift bin centred at 0.65 to 86 percent at the mean redshift of 1.1. Furthermore, unobscured starbursts dominate our overall source list in all the redshift bins.
This could also be a selection effect. Fig. \ref{fig:gal_type_contr} does not necessarily imply that extinction changes with redshift, but the trend seen may simply reflect the fact that at higher redshifts we are only probing the most UV-luminous sources, whereas the sources with extinction contribute most at the lower UV luminosities, as seen in Fig. \ref{fig:lf_kcorr_test} and associated discussion in the previous section.

\subsection{Luminosity functions}
\label{sec:6.2}
\begin{figure*}
    \centering
    \hspace*{-0.2cm}\includegraphics[width=0.87\textwidth]{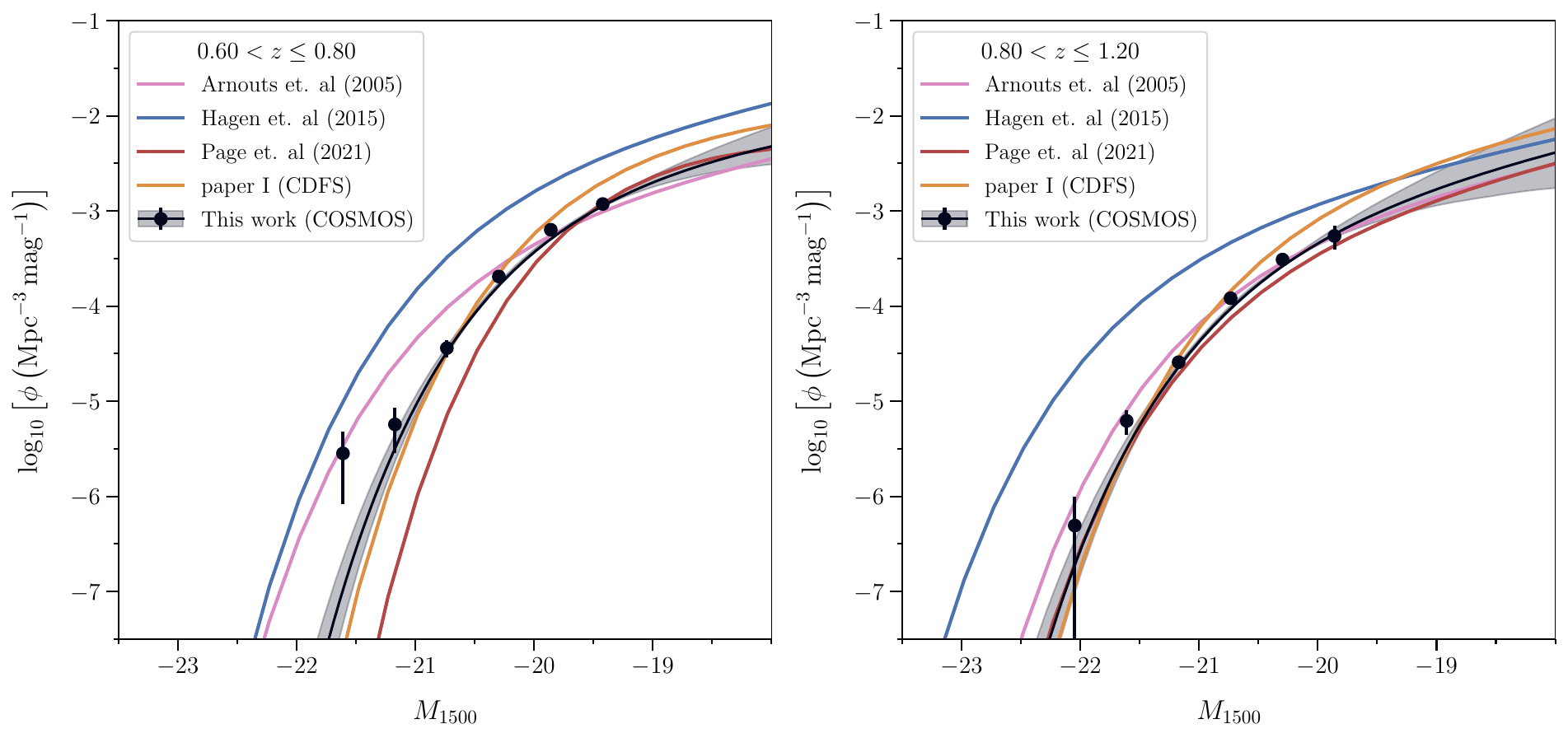}
    \caption{UV luminosity function of galaxies in the redshift
    intervals $0.6 \leq z < 0.8$ in the left panel and $0.8 \leq z < 1.2$
    in the right panel as a function of the 1500 {\AA}  magnitude.
    Data points show the binned number densities measured using the \citet{2000MNRAS.311..433P} method. The solid black line is our best-fitting Schechter function derived from the CDFS field, as described in Section \ref{sec:4.2}.
    We obtain this curve from the median value of the posterior distribution
    of Schechter function parameters.
    The grey shaded area around the best-fit Schechter function
    represents the $1 \sigma$ (68.26 percent) uncertainties.
    Solid blue, red, and purple lines are the Schechter functions obtained by \citet{2005ApJ...619L..43A}, \citet{2015ApJ...808..178H}
    and \citet{2021MNRAS.506..473P}.}
    \label{fig:lf}
\end{figure*}

\begin{figure*}
    \centering
    \hspace*{-0.2cm}\includegraphics[width=1.2\columnwidth]{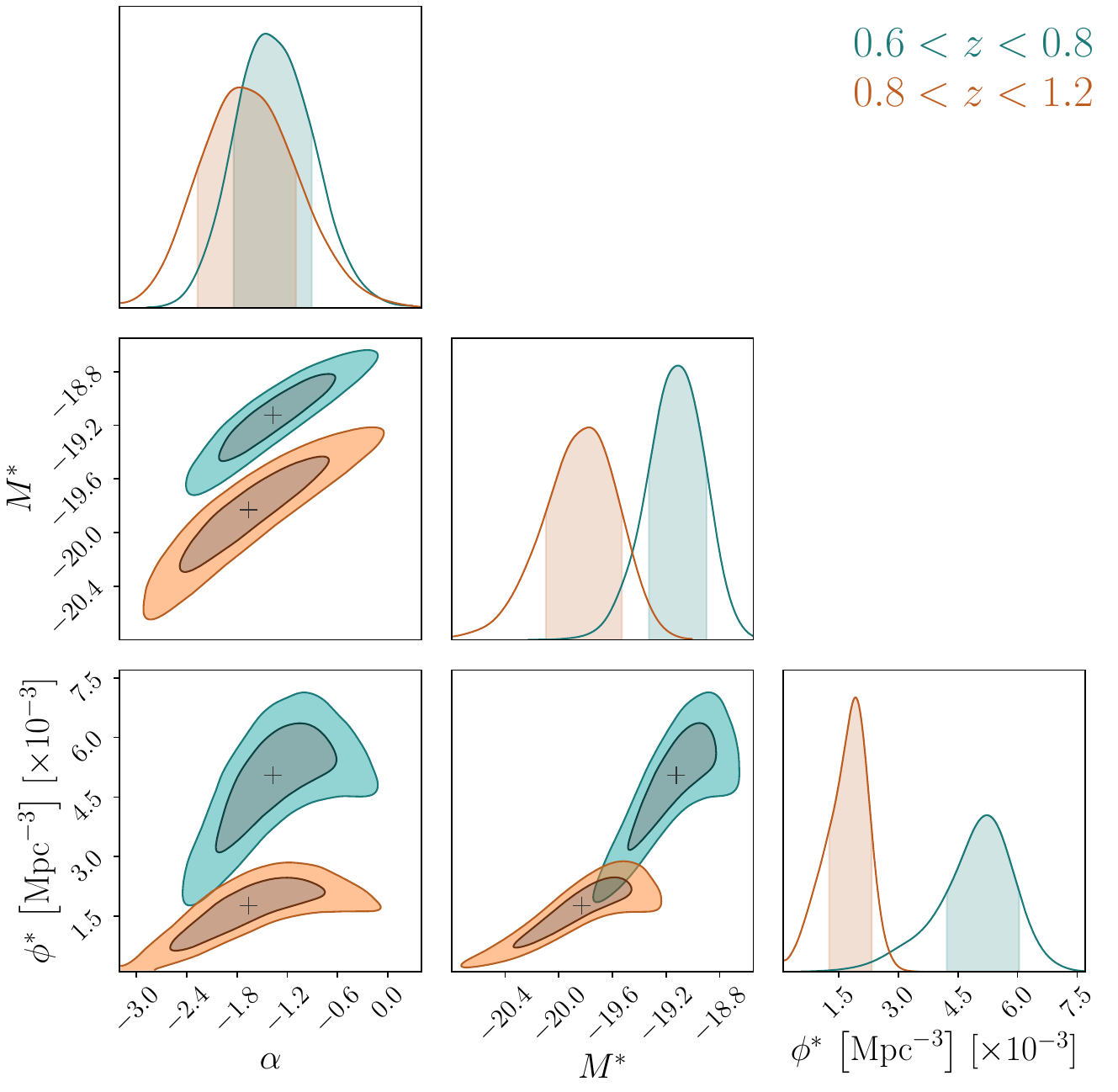}
    \caption{
    This figure represents the marginalized one-dimensional (along the diagonal)
    and two-dimensional (off-diagonal) posterior distributions of Schechter
    function parameters $\alpha$, $M^{*}$ and  and $\phi^{*}$. 
    The redshift bin $0.6 \leq z < 0.8$ is represented by blue, and the redshift bin $0.8 \leq z < 1.2$ is shown in red colour. 
    The shaded region in the dark and light-coloured areas in the off-diagonal 
    part of the plot corresponds respectively to 68 and 95 percent confidence 
    intervals for LF parameters.
    The black `+' symbols represent the median values for $\alpha$, $M^{*}$
    and $\phi^{*}$. The shaded region in the diagonal plots represents one
    dimensional 68 percent confidence region.}
    \label{fig:ci}
\end{figure*}

The LF obtained from the UVW1 sample are plotted in Fig. \ref{fig:lf}, along with the best-fit 
Schechter function models from \citet{2005ApJ...619L..43A,2015ApJ...808..178H,2021MNRAS.506..473P} 
and \citetalias{2022MNRAS.511.4882S} at the same redshifts.
We show in Fig. \ref{fig:ci}, the one- and two-dimensional posterior distributions
for the LF parameters, obtained by MCMC simulations.
The dark and light-shaded regions show 68 and 95 percent confidence regions for the Schechter function parameters.
The best-fit values obtained using the maximum likelihood method presented in Section
\ref{sec:4.2} are listed in Table \ref{tab:lf_bins}.

\begin{figure*}
    \centering
    \hspace*{-1.37cm}\includegraphics[width=0.90\columnwidth]{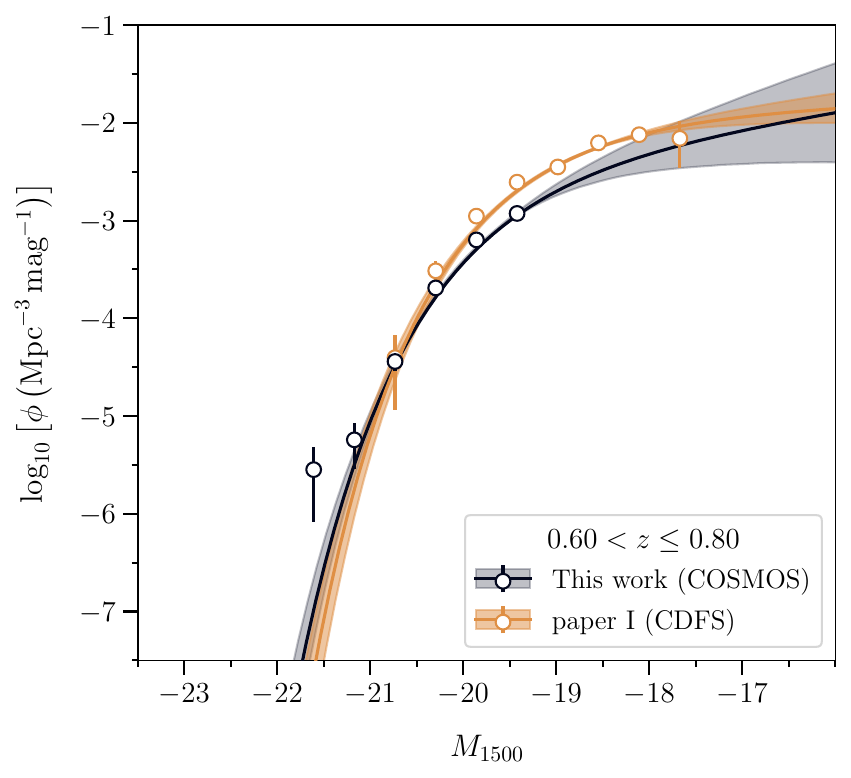}
    \hspace*{1.47cm}\includegraphics[width=0.90\columnwidth]{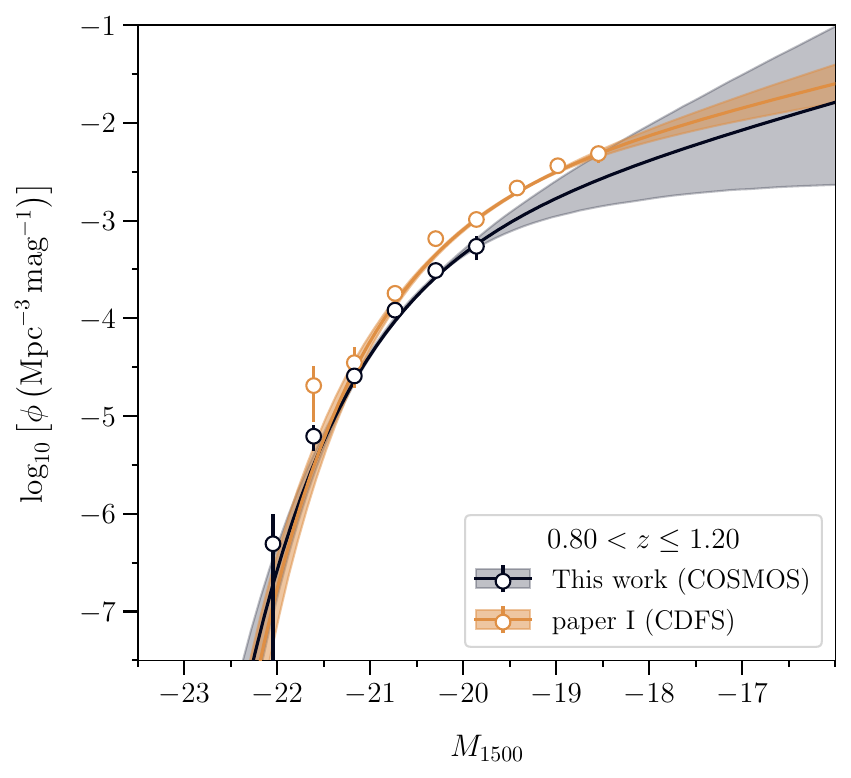}
    \hspace*{-0.5cm}\includegraphics[width=0.95\columnwidth]{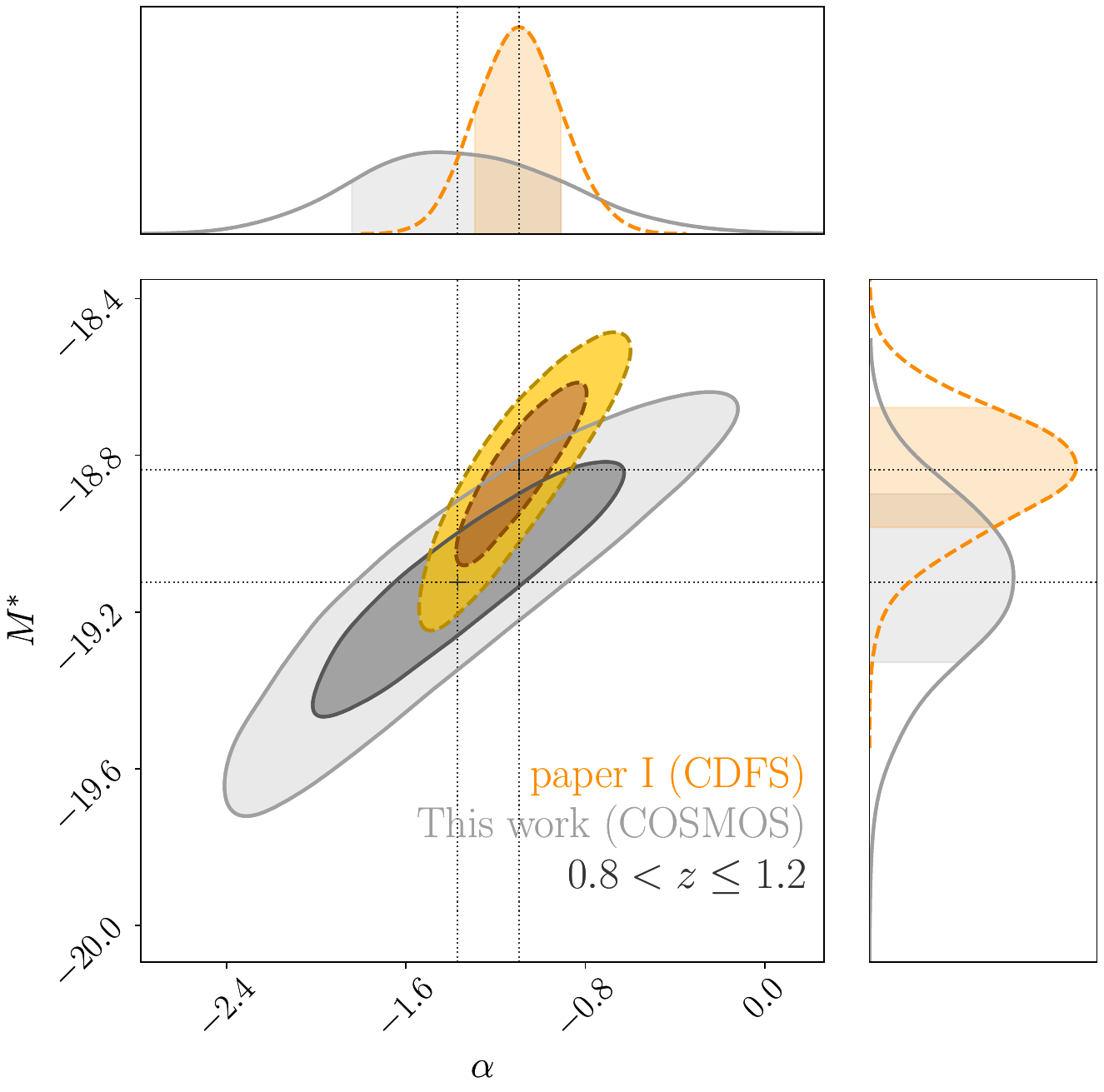}
    \hspace*{1.1cm}\includegraphics[width=0.95\columnwidth]{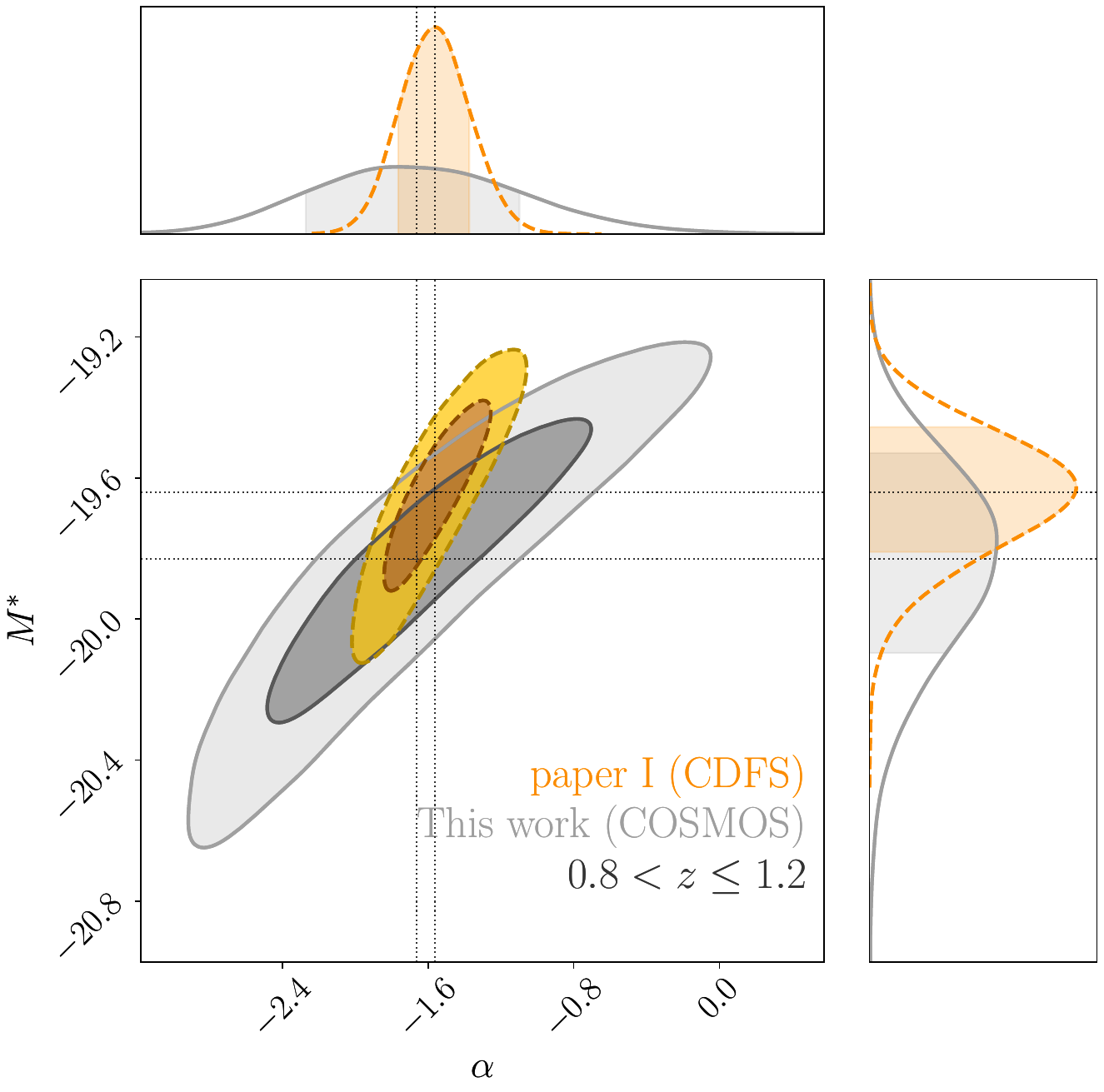}
    \caption{The comparison of the UV galaxy luminosity function of the 
    CDFS (yellow) and COSMOS (black/grey) in the redshift interval $0.6 - 0.8$ (left) and $0.8 - 1.2$ (right). 
    Upper left and right panels: The data points show the binned number densities and the solid lines are maximum likelihood fitted Schechter functions. The
    shaded regions represent the $1\sigma$ uncertainties respectively for 
    redshift 
    bins $0.6 - 0.8$ and $0.8 - 1.2$.
    Lower left and right panels: The one and two-dimensional marginalised distributions in
    the $M^*-\alpha$ space for the COSMOS (grey) and CDFS (yellow), for lower 
    and higher redshift bins. The dark and light-shaded regions represent the $1\sigma$ and $2\sigma$ confidence
    regions, and the `+' symbols denote the median value of the parameters.}
    \label{fig:lf68comp1}
\end{figure*}

\begin{figure*}
    \centering
    \hspace*{-0.1cm}\includegraphics[width=0.85\textwidth]{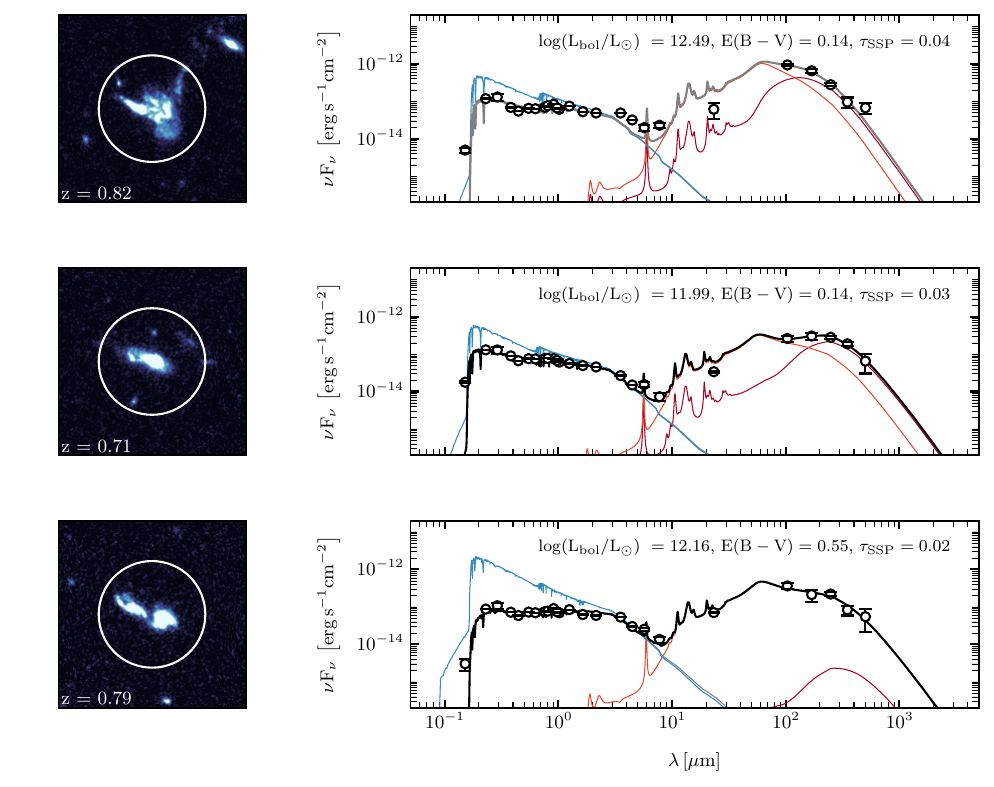}
    \caption{We plot here the SEDs and \textit{HST} ACS stamps of the brightest star-forming galaxies in our sample. The \textit{top} row shows the most luminous (UVW1 AB 21.15 mag) source in the redshift bin $0.8-1.2$; the \textit{middle} and the \textit{bottom} rows show two (UVW1 AB mag = 21.16, 21.35) of the most luminous objects in the redshift range $0.6-0.8$.
    These sources constitute the brightest magnitude bin of their respective redshift bins.
    The panels on the left show the \textit{HST} ACS stamps of these sources and their redshifts, and the panels on the right-hand side panels show the SEDs. The size 
    of the stamps is $10 \times 10$ sq arcseconds. A circular aperture (3 arcsec radius) used for UVW1 photometry COSMOS image is also shown in white colour. 
    The panels on the right-hand side show the SEDs.
    The total SEDs, shown in black solid
    lines represent the linear sum of the stellar models generated 
    using \citet{2003MNRAS.344.1000B}
    and the two infra-red models from \citet{2007ApJ...657..810D}, one for a 
    constant (orange) and
    another for a variable (maroon) radiation field. The blue line represents the stellar model without reddening (i.e. $E(B-V)=0$).
    In each SED panel, we also show the fitted age of the simple stellar population (in Gyr), the colour excess, and the bolometric luminosity.}
    \label{fig:seds}
\end{figure*}

\begin{table}
  \setlength{\tabcolsep}{10pt}
  \centering
  \caption{Derived Schechter function parameters of the galaxy UV 
  LF from their respective posterior distributions at both redshift bins. 
  Errors indicate $1 \sigma$ (68.26 percent) uncertainties.}
  \label{tab:lf_bins}
  \begin{tabular}{ccrc}
    \hline\hline
    \noalign{\vskip 0.5mm}
    $z$ &
    \multicolumn{1}{c}{$\phi^* / 10^{-3}$} &
    \multicolumn{1}{c}{$M^*$} &
    \multicolumn{1}{c}{$\alpha$} \\
    &
    \multicolumn{1}{c}{(Mpc$^{-3}$)}
    &
    & \\
    \hline
    \noalign{\vskip 0.5mm}
    $0.6 - 0.8$ & $5.05^{+0.75}_{-1.07}$ & $-19.12_{-0.22}^{+0.20}$ & $-1.37_{-0.43}^{+0.48}$ \\
    \noalign{\vskip 0.5mm}
    $0.8 - 1.2$ & $1.76^{+0.43}_{-0.64}$ & $-19.83_{-0.29}^{+0.26}$ & $-1.66_{-0.55}^{+0.59}$ \\
    \noalign{\vskip 0.5mm}
    \hline
  \end{tabular}\\
  \label{tab:par_f}
\end{table}

\begin{table}
  \setlength{\tabcolsep}{20pt}
  \centering
  \caption{Cosmic variance errors on normalisation calculated using two 
  alternate methods explained in Section \ref{sec:4.5}. We tabulate the 
  average stellar masses (second column) in both redshift bins (column 1).
  The last column shows $1 \sigma$ fractional errors in normalisation calculated using 
  \citet{2008ApJ...676..767T} (I) and \citet{2010ApJ...710..903M} (II).}
  \label{tab:cv}
  \begin{tabular}{cccc}
    \hline\hline
    \noalign{\vskip 0.5mm}
    $z$ &
    \multicolumn{1}{c}{$M_*/ 10^{10}$} &
    \multicolumn{2}{c}{$\Delta \phi^{*} / \phi^{*} (1 \sigma)$} \\
    \cline{3-4}
    \noalign{\vskip 0.5mm}
    &
    \multicolumn{1}{c}{(${M_\odot}$)} &
    I &
    II \\
    \noalign{\vskip 0.5mm}
    \hline
    \noalign{\vskip 0.5mm}
    $0.7$     & 1.43        & 0.10  & 0.11  \\
    $1.0$     & 1.73        & 0.08  & 0.07  \\
    \hline
  \end{tabular}\\
\end{table}

We can see in Fig. \ref{fig:lf68comp1} (upper panels) that the wide area
COSMOS survey extends the bright end of the LF by almost a magnitude
in the redshift bin $0.6 - 0.8$ and half a magnitude at $0.8 - 1.2$ compared to the CDFS LFs.
As a consequence of this, we can probe space densities of an order of magnitude lower
in COSMOS than in CDFS, of order $10^{-6}$ Mpc$^{-3}$ Mag$^{-1}$.
We notice the difference in the normalisations of the LF between the two redshift bins. These differences can be attributed to different space densities of galaxies in the CDFS and COSMOS fields (cosmic variance).

Our measurements are fairly consistent with the Schechter function shape in both redshift bins.
We can see this in the top right panel of Fig. \ref{fig:lf68comp1}, where the 
observed binned LF
follows the shape of the modelled Schechter function. A similar behaviour is
observed in the redshift bin $0.6-0.8$ (top left of Fig. \ref{fig:lf68comp1}), except at the brightest absolute magnitude
bin, in which the binned LF appears to considerably exceed the Schechter function model. Therefore, we examine this bin in more detail to determine whether the data in this bin pose a serious challenge to the Schechter function model. 

In this brightest absolute magnitude bin, we calculated from the Schechter function model that the expected number of sources in our survey is 0.12, whereas the number of galaxies observed is 2. From Poisson statistics, the probability of observing two or more galaxies when the expected number is 0.12 is $7.1\times 10^{-3}$, a discrepancy that is a little less than 3$\sigma$, but still a cause for concern.

Given this observed discrepancy, we have looked in more detail at the sources in the brightest absolute magnitude bins in both magnitude ranges, and in particular at the possibility that they are contaminated by AGN.

Some potential AGN candidates
might be missed by the X-ray detection because the
X-ray observations do not reach $10^{42}\,\,\mathrm{ergs\,s}^{-1} \mathrm{cm}^{-2}$ in all three bands ($0.5 - 10$, 
$0.5 - 2$ and $2 - 10$ KeV) up to redshift 1.2 (refer to Section \ref{sec:2.4.3}). Therefore there may be AGNs which are hiding below the X-ray detection limit and are present in our final UVW1 source-list. In the construction of the LF, there is a significant potential for AGN contamination at the high luminosities, where the number densities are very low. Therefore, caution must be taken, especially at the bright end of the LF.

The first step we take to address this issue is to examine the mid-IR properties  
of these brightest sources. There are two sources in the brightest absolute magnitude bin in the $0.6-0.8$ redshift range and a single source in the brightest absolute magnitude bin between redshifts 0.8 and 1.2. For these sources, we obtained the magnitudes in the W1 ($3.4 \micron$) and W2 ($4.6 \micron$) passbands of the \textit{Wide-field Infrared Survey Explorer} \citep[\textit{WISE};][]{2010AJ....140.1868W} and the 3.6 and 4.5 $\micron$ passbands of the \textit{Spitzer} Infrared Array Camera \citep[IRAC;][]{2004ApJS..154...10F}.
We checked the sources against the mid-IR AGN identification criterion set out for \textit{WISE} colours by
\citet{2012ApJ...753...30S} and \citet[][]{2013ApJ...772...26A}. Similarly, for IRAC colours, we used the prescription from \citet{2020A&A...640A..68I}.
None of these sources
satisfies the conditions outlined for the \textit{WISE} and IRAC colours, by the above-mentioned works.
Thus, none of the sources populating the highest luminosity bins in our LFs shows evidence of a substantial energetic contribution from an AGN.

To look further at these three sources, we examine their SEDs. We also examined their morphology using the COSMOS image from the \textit{Hubble Space Telescope} (\textit{HST}) Advanced Camera for Surveys (ACS) which employed the F814W filter \citep[][]{2007ApJS..172..196K}. The SEDs along with the 10$\times$10 arcsecond
postage stamp images are shown in Fig. \ref{fig:seds}.

As seen in Fig. \ref{fig:seds} these luminous galaxies are detected from the far-UV to submillimeter wavelengths. 
We fit SEDs containing stellar and dust components (Section \ref{sec:4.6}).
The fits suggest that all three objects have young stellar populations and
significant amounts of dust.
The reddening derived for these sources is at most $E(B-V)=0.55$, and is degenerate to some extent with the contribution of older stars which were not taken into account in the SED fitting, and so the actual reddening may be less than indicated by your simple SSP fitting.
We obtain their total IR luminosity by integrating the dust components. 
These sources cover a range of IR luminosities from $\num{9.88e11}{L_\odot}$ to
$\num{5.04e12}{L_\odot}$.
Therefore, these most powerful UV-selected galaxies in the COSMOS field correspond to
(U)LIRGs. These systems are defined by their total infrared luminosity as $-$
LIRGs; $10^{11}{\mathrm{L}_\odot} < L_\mathrm{IR}(8-1000\micron) < 10^{12}{\mathrm{L}_\odot}$ and 
ULIRGs; $10^{12}{\mathrm{L}_\odot} < L_\mathrm{IR}(8-1000\micron) < 10^{13}{\mathrm{L}_\odot}$ \citep[][]{1996ARA&A..34..749S,1998ApJ...498..579G}.
However, none of these systems are as luminous as the most powerful IR galaxies 
in the redshift range under consideration. 
\citet{2010ApJ...709..572K} in their 70 $\micron$ selected catalogue find sources 
as bright as $\sim \num{8e12}{L_\odot}$ in a redshift range similar to ours.
One of the reasons we do not find any such sources could be that the most powerful
IR galaxies are bright AGNs \citep[][]{1996ARA&A..34..749S,2010ApJ...709..572K,2011MNRAS.414.1903G,2021MNRAS.503.3992S}, which we remove from our analysis. The other reason could be the UV selection, as the brightest IR galaxies could be too obscured to be seen strongly at the UV wavelengths.

Examination of the HST images indicates that the morphology of at least two out of the three sources constitute mergers in various stages.
The brightest source in the sample (top panel in Fig. \ref{fig:seds}) seems to be in an advanced stage of a merger. Examining the spectral energy distribution (SED), it becomes apparent that the UV and 24 $\mu$m data points do not show a good fit. 
There could be multiple reasons for this. There could be two distinct stellar populations, corresponding to the pre-merger sources, making it challenging to fit with a single stellar population model.
Another possibility is that the disturbed morphology of the merger is giving rise to a distribution of obscuration across different parts of the galaxy. This distribution could be more heterogeneous than what our current modelling can effectively account for.
In contrast, the source in the bottom panel of Fig. \ref{fig:seds}, falling within the redshift range of 0.6-0.8, seems to represent an early-stage merger, with two discrete luminous galaxies within the XMM-OM UVW1 photometry aperture. It is plausible that they possess stellar populations of similar ages, allowing a single stellar population model to reasonably fit their photometric data. However, it is worth noting that the UV data points still slightly deviate from the fitted model.
The important point here is the identification of this UVW1 source as two discrete galaxies, which offers a solution to the discrepancy between the observed and model-predicted number of galaxies in the brightest absolute magnitude bin at $0.6<z<0.8$. If the UV emission of the galaxies were measured separately, their UV absolute magnitudes may well place them in a different bin of the luminosity function. Thus, the number of galaxies observed in this most luminous bin may only be one, rather than two, in which case the discrepancy is no longer significant, and we no longer observe a deviation from the Schechter function shape at the bright end.

\subsection{Comparison with literature}
\label{sec:6.3}
We compare our LFs with those found in previous studies. We begin by
comparing our results to those of \citetalias{2022MNRAS.511.4882S}. In terms of $M^*$ and $\alpha$
the bottom of Fig. \ref{fig:lf68comp1} 
shows that the contours of our study and those of \citetalias{2022MNRAS.511.4882S} overlap. The
best-fit values of the faint-end slopes
and the characteristic magnitudes are within
$1\sigma$ and $2\sigma$ of each other, respectively, for the redshift
bin centred at $z=0.7$ and $1\sigma$ for $z=1.0$. 
The contours for the CDFS are smaller, which shows that the
LF parameters are better constrained than those of the COSMOS field.
We attribute this to the depth of the CDFS UVW1 survey, which probes fainter absolute magnitudes and hence covers the transition between exponential and
power law parts of the Schechter function as well as the faint end slope.
The large errors in the faint-end slope drive the errors in the characteristic magnitude
due to the strong correlation between the two quantities. Fixing the faint-end
slope in this study to the values obtained in the CDFS, we obtain $M_*$ of $-19.06 \pm 0.07$ and $-19.79 \pm 0.08$ at redshifts of 0.7 and 1.0, respectively.

In general, our values of the UV LF parameters are in agreement with the
findings reported by \citet{2005ApJ...619L..43A}. The faint-end slope estimates of
$\alpha = -1.60 \pm 0.26$ and $-1.63 \pm 0.45$ from their study are within $1 \sigma$ of
our estimate of $\alpha = -1.37_{-0.43}^{+0.48}$ and $-1.66_{-0.55}^{+0.59}$, for redshift bins at $z=0.7$ and $z=1.0$
respectively.
However, we do see a slight deviation from their LF, in the redshift range
$0.6 < z < 0.8$ (left panel of Fig. \ref{fig:lf}).
At bright absolute magnitudes ($-21.5<M_{1500}<-20.5$) their model LF curve (pink) lies significantly above our measurements.
In parameter terms, this discrepancy corresponds to the fainter best-fit characteristic magnitude we obtain in this redshift range ($M^{*} = -19.12_{-0.22}^{+0.20}$) compared to that found by \citet{2005ApJ...619L..43A} ($M^{*} = -19.84 \pm0.40$).
In the other redshift bin (that is, $0.8 < z <1.2$), the best-fit model curve of \citet{2005ApJ...619L..43A} lies close to our LF measurements, and our measurement of $M^{*}$ agrees with that obtained by \citet{2005ApJ...619L..43A} within 1$\sigma$. 
In comparison of $M^{*}$ with \citet{2015ApJ...808..178H}, we find that
our values are fainter by at least $3\sigma$ in both redshift bins.
Their model LF curve (blue) in the lower redshift bin is above our measurements throughout the magnitude range considered in this study. The differences in
the higher redshift bin becomes more severe as the absolute magnitude
brightens. These differences can be attributed to our careful elimination of the bright AGN, compared to \citet{2015ApJ...808..178H} whose results might be affected by AGN contamination. This is discussed in detail in the appendix of \citetalias{2022MNRAS.511.4882S}.

\begin{figure}
    \centering
    \hspace*{-0.3cm}\includegraphics[width=\columnwidth]{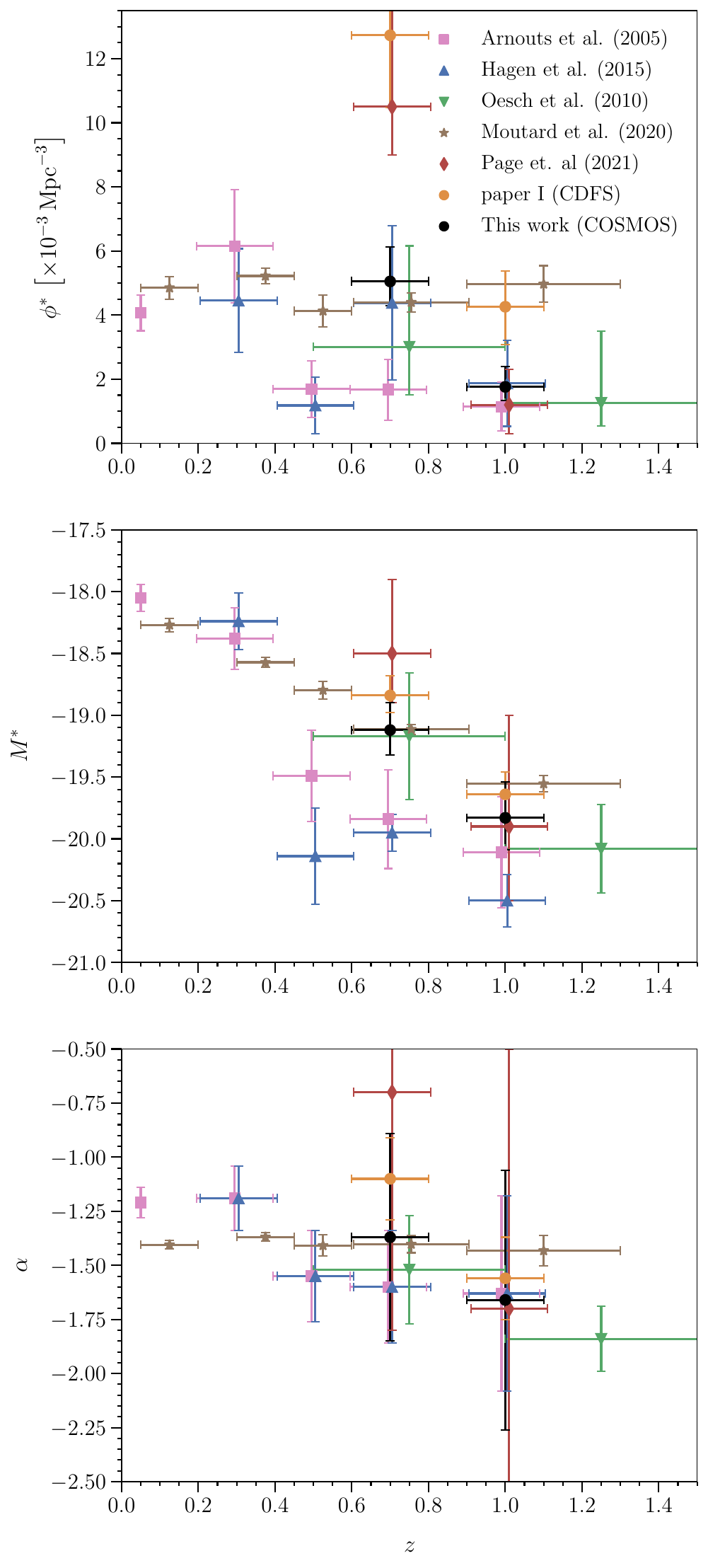}
    \caption{The parameters $\alpha$ and $M^{*}$ and $\phi*$ of the 
    Schechter function as defined in Equation \ref{eqn:phi} and tabulated
    in Table \ref{tab:par_f}.
    The values estimated from this paper are in black. Other colours are used 
    for values estimated by other studies of \citet{2005ApJ...619L..43A,2015ApJ...808..178H,2010ApJ...725L.150O,2020MNRAS.494.1894M,2021MNRAS.506..473P} and \citetalias{2022MNRAS.511.4882S}.
    Only those works that use direct UV observations to calculate their LF, are shown in this Figure.}
    Different panels from top to bottom represent the normalisation $\phi*$, characteristic magnitude $M^{*}$ and faint-end slope $\alpha$
    as a function of redshift.
    The horizontal and vertical error bars represent the width of the redshift bin and 1 $\sigma$ (68.26 percent) uncertainties
    respectively. For clarity, the values of the parameters at the same redshifts are slightly shifted in redshift.
    \label{fig:comp_par}
\end{figure}

Compared to other recent work, we find that our results for $\alpha$ are
in agreement with \cite{2021MNRAS.506..473P}.
The work of \citet{2010ApJ...725L.150O} (for $z=0.75$), who have
used the deepest data to date to calculate the LF, obtain parameters
which are in good agreement with our results.
Their result for $\alpha$ and the break magnitude $M^{*}$ agrees with our
values with deviations well within $1 \sigma$.

The two important studies that calculate galaxy LF in the redshift range $0.6-1.2$ using ground-based instruments are \citet{refId0} and \citet{2020MNRAS.494.1894M}. They derived their estimates by extrapolating their observations taken at
longer wavelengths to the UV regime using SED templates. For completion, we
do compare our results to these works and add the data points from these studies to the luminosity density plot but note that comparison of these works with
the studies dealing with direct UV surveys should be taken with caution.  
Compared to \citet{refId0}, we find that our estimates of the faint-end slope
at redshift 0.7 are within $1\sigma$ of theirs. In the redshift bin, our value for $\alpha$ is well within the $2\sigma$ level from their values
in redshift bins centred at 0.9 and 1.1.
However, we do find a larger discrepancy when we look at the characteristic
magnitudes. Compared to our estimates at redshift 0.7, their value is fainter by
at least $4\sigma$. For their higher redshift bins at 0.9 and 1.1, we see that the
difference in the $M^*$ values is $4\sigma$ and $2\sigma$ significant, compared to our estimate at redshift 1.0.
With respect to \citet{2020MNRAS.494.1894M} ($z=0.75$), our values agree well.

We do not place much emphasis on the comparison of the normalisation values
$\phi^*$ obtained in this work with
any of the previous works and/or \citetalias{2022MNRAS.511.4882S}. These differences in $\phi^*$ 
are expected due to cosmic variance
\citep[][]{2008ApJ...676..767T,2010ApJ...710..903M}, between different parts of the Universe explored by different studies.
Nevertheless, for completion, we plot the
estimates for $\phi^*$ in Fig. \ref{fig:comp_par} (top panel) from
this work and some other studies.
We would like to remind the reader
about the additional errors due to cosmic variance (Section \ref{sec:4.5})
to be considered, while comparing the resulting value with other works at similar redshifts.
Among previous works based on direct observation of the 1500 {\AA} 
radiation only \citetalias{2022MNRAS.511.4882S}, \citet{2021MNRAS.506..473P} and \citet{2015ApJ...808..178H}
estimate the uncertainties in their measurements due to cosmic variance, so we compare our results for normalisation only with these studies.
Our estimates for $\phi^*$ are in very good agreement with \citet{2015ApJ...808..178H} in both redshift bins. 
Regarding \citet{2021MNRAS.506..473P}, our values agree at $2\sigma$ and
$1\sigma$ at redshifts 0.7 and 1.0. 
It should be mentioned, however, that due to the small sample size, the values obtained have large statistical errors. Compared to \citetalias{2022MNRAS.511.4882S},
we notice smaller values of $\phi^*$ in both redshift bins. These differences, which are significant at $\sim 2\sigma$, can
be attributed to known galaxy clusters in the CDFS explored in \citetalias{2022MNRAS.511.4882S}.
Due to the larger sky area of the COSMOS image, here we have managed to get better constraints on both the normalisation and the cosmic variance.
We want to remark here that more independent UV surveys can further help with properly constraining the normalisation of the galaxy LF.

\begin{figure}
  \hspace*{-0.3cm}\centering
  \includegraphics [width=\columnwidth]{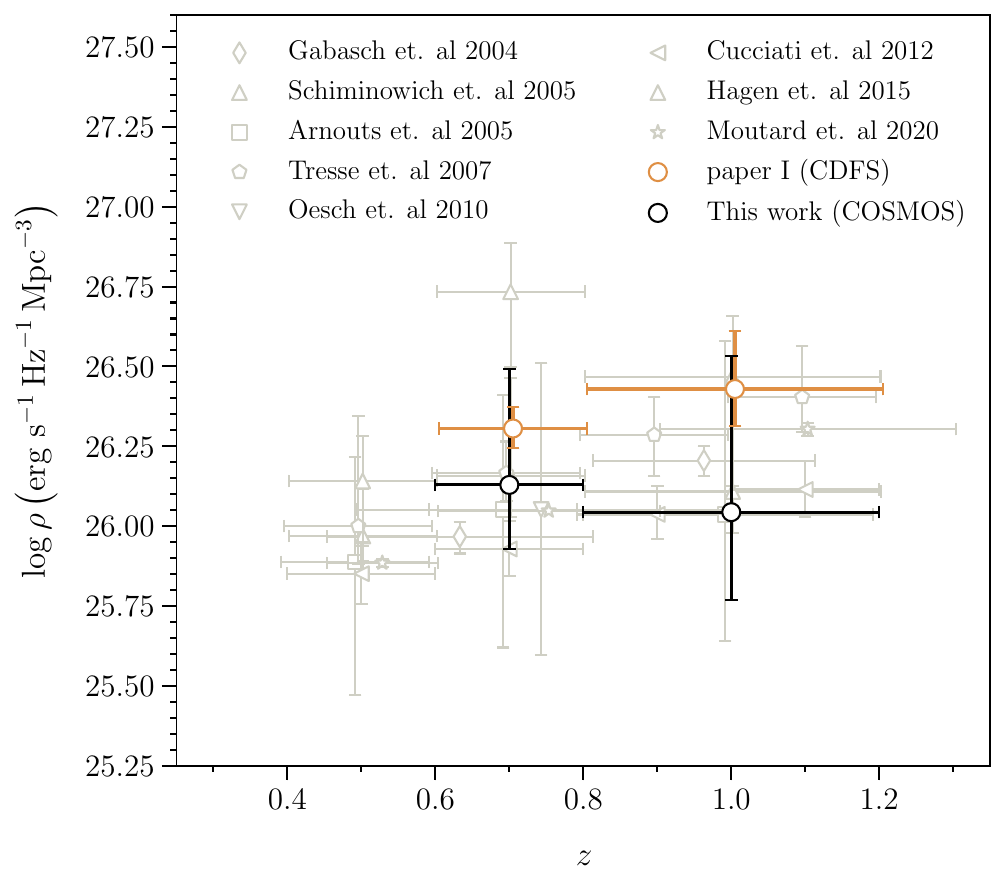}
  \caption{The luminosity density calculated using Equation \ref{eqn:j2}. The grey data points are the observed luminosity densities from different past studies. Our estimates for the present work (COSMOS) are shown in black colour and those for \citetalias{2022MNRAS.511.4882S} (CDFS) are in yellow. The vertical error bars represent the $1 \sigma$ (68.26 percent) uncertainties. The
  horizontal ones are redshift bin sizes. The data points at the same 
  redshifts are slightly shifted for clarity.}
  \label{fig:lum_dens}
\end{figure}

The luminosity density (LD) in the two redshift bins is calculated as shown in
Section \ref{sec:4.3}. Our values fall within the error margins of previous work as illustrated in Fig. \ref{fig:lum_dens}. Given the large area of the COSMOS UVW1 image, the impact of cosmic variance on our results is minimal compared to the statistical uncertainties.
Only about 10 and 15 percent of the error budget is contributed by cosmic variance.
The remaining uncertainty in the value of the LD is driven by the large errors in the faint-end slope of the LF. 
Comparing the COSMOS and CDFS datasets, it becomes evident that the errors associated with LD measurements in the COSMOS field are more significant. This discrepancy can be attributed to both substantial statistical errors in the faint-end slope and a smaller sample size in COSMOS compared to CDFS.
Given these large errors in the LD for the COSMOS field, the CDFS values are within the $1 \sigma$ distance of the values for redshifts 0.7 and 1.0. The overall luminosity density values are marginally lower in the COSMOS dataset, potentially due to the higher galaxy number density observed in the CDFS dataset. This distinction is also evident in the top panels of Fig. \ref{fig:lf68comp1}, where the LFs in COSMOS exhibit higher normalisation in both redshift bins.

\begin{table}
  \setlength{\tabcolsep}{20pt}
  \centering
  \caption{Luminosity density as a function of redshift. 
  Errors indicate $1 \sigma$ uncertainties, which include $1 \sigma$
  relative error from cosmic variance.}
  \begin{tabular}{ccc}
    \hline\hline
    \noalign{\vskip 0.5mm}
    $z$ &
    \multicolumn{2}{c}{$\rho / 10^{26}$} \\
    &
    \multicolumn{2}{c}{$(\mathrm{erg}\,\mathrm{s}^{-1} \mathrm{Hz}^{-1} \mathrm{Mpc}^{-3})$} \\
    \cline{2-3}
    \noalign{\vskip 0.5mm}
    &
    This work &
    \citetalias{2022MNRAS.511.4882S}$^{a}$ \\
    \noalign{\vskip 0.5mm}
    \hline
    \noalign{\vskip 0.5mm}
    $0.6 - 0.8$ &  $1.34_{-0.49}^{+1.48}$  & $2.02_{-0.27}^{+0.33}$  \\
    \noalign{\vskip 0.5mm}
    $0.8 - 1.2$ &  $1.10_{-0.51}^{+1.66}$  & $2.69_{-0.63}^{+1.43}$  \\
    \hline
  \end{tabular}\\
   \begin{minipage}{0.92\columnwidth}
      \textsuperscript{$a$}{The error bars on luminosity density in Table \href{https://academic.oup.com/view-large/389669641}{5} of \citetalias{2022MNRAS.511.4882S} do not include the cosmic variance.}
  \end{minipage}
  \label{tab:lumden}
\end{table}

\subsection{Evolution of the LF and LD}
\label{sec:6.4}

Between the two redshift bins, we do not observe any evolution of the
faint-end slope. This is in line with \citetalias{2022MNRAS.511.4882S} and other previous work.
However, we do see a $\sim 2\sigma$ variation in the characteristic
magnitude. As we move from redshift 0.7 to 1.0, $M^*$ brightens by
0.7 mag. This is very close to the 0.8 mag evolution seen in \citetalias{2022MNRAS.511.4882S} and a
similar change in $M^*$ is observed by \citet{2015ApJ...808..178H}.
This brightening of the characteristic magnitude supports the currently
accepted notion of the star formation history of the Universe, i.e.
the star formation rate decreases as we move from redshifts of around 2 to smaller values.
As mentioned earlier, differences in normalisation are expected
due to cosmic variance. We note here, that these errors need to be considered
not only for comparison with other studies but also for looking at the potential evolution
of normalisation with redshift.
As mentioned earlier, due to the large area of the COSMOS UVW1 image, 
the uncertainties
in normalisation due to cosmic variance (\num{5.4e-2} $\mathrm{Mpc}^{-3}$ and 
\num{1.4e-2} $\mathrm{Mpc}^{-3}$ for 
$z = 0.7$ and $z = 1.0$ respectively) are smaller than the statistical uncertainties. Nevertheless, we include these errors due to cosmic variance in the quadrature to statistical errors.
We see an evolution in the normalisation of the UV LF as it goes down 
with more than $2\sigma$ significance as we move from the lower to the higher redshift bin.

For the case of the luminosity density (LD), we do not see any significant evolution between redshifts 0.7 to 1.0.

\section{Conclusion}
\label{sec:7}

We use the wide area COSMOS imaging survey data, specifically taken through the UVW1 filter 
on XMM-OM, to calculate the UV LF of galaxies within a redshift range spanning from 0.6 to 1.2. Using this survey, we aim to expand our analysis to brighter magnitudes, which complements our earlier work in \citetalias{2022MNRAS.511.4882S}, where we determined the UV LF using deep survey data from the CDFS, primarily to constrain the faint end of the LF.

The binned UV galaxy LF is estimated in two redshift intervals: 0.6 < z < 0.8 and 0.8 < z < 1.2, employing the method outlined in \citet{2000MNRAS.311..433P}. Notably, the COSMOS imaging data pushes the rest-frame $M_{1500}$ magnitudes to higher values, approximately $-21.5$ for the redshift bin 0.6 < z < 0.8, and roughly $-22$ for the redshift bin 0.8 < z < 1.2. This enhancement enables us to impose more precise constraints on the characteristic magnitude ($M^*$) in our analysis.

We analyze the reddening properties of the UVW1 COSMOS sample by examining the optical $B-I$ colours of the sources. Our findings indicate that the sample predominantly consists of starburst galaxies, with a major contribution of galaxies exhibiting low levels of reddening. 
This substantiates our use of low-extinction starburst templates for K-corrections in our study.
When we examine the LF of two subgroups characterised by low and moderate levels of reddening, it becomes apparent that sources with moderate reddening become increasingly significant in the lower luminosity bins of the LF. Interestingly, the combined LF of these low and moderately reddened subgroups matches the LF of the entire sample, further reinforcing the dominance of low-extinction starbursts in our dataset. Additionally, we observe a growing proportion of low reddening sources as the redshift increases.

We fit the Schechter function to the data using the maximum likelihood to estimate the LF parameters (the faint-end slope, characteristic magnitude and the normalisation).
We compare the binned LF shape to the Schechter model. The luminosity function seems to be well described by the Schechter function shape.
There is no significant evolution of $\alpha$ between the two redshift bins. This observation aligns with previous studies that have consistently found a nearly constant faint-end slope within the redshift range of our interest.
For individual redshift bins, the value of $\alpha$ falls within the $1\sigma$ confidence intervals of values obtained in the \citetalias{2022MNRAS.511.4882S}, as well as other studies within this redshift range using direct UV measurements.

Turning to the characteristic magnitude $M_{*}$, our derived values are 0.5 to 1 mag fainter than some previous studies. Between the redshift bins under consideration, $M_{*}$ evolves by $\simeq 0.7$ mag
with a $\sim 2\sigma$ significance. A similar evolution of approximately 0.8 magnitudes in $M_{*}$ between these redshift bins was also reported in \citetalias{2022MNRAS.511.4882S}. When we compare our estimates at a specific redshift to prior works (again based on direct UV surveys), our results fall within the $2\sigma$ confidence intervals, indicating good agreement.

As expected, our calculations for luminosity density yield values that are relatively smaller than those obtained in the CDFS. However, the differences are not statistically significant, remaining below the $2\sigma$ threshold for both redshifts. Between the two redshift bins considered here, the change in luminosity density is not significant enough to infer evolution.

In this study, we focus special attention on the bright sources. First, we examine the most luminous sources, found within the brightest magnitude bins of the LF at both redshifts, to detect any potential contamination from active galactic nuclei (AGN). 
Subsequently, we characterise these objects through an SED analysis and morphological examination. Notably, some of the most luminous UV sources in the COSMOS field appear to be mergers at various stages of evolution. Their SEDs and integrated IR luminosities suggest that these galaxies belong to the (U)LIRG class of galaxies, albeit not the most powerful representatives. We attribute this observation to our exclusion of bright AGNs from the analysis or the possibility that the most luminous galaxies are more heavily obscured than our UV-selected galaxies.

\section{Acknowledgements}
\label{sec:8}
This research uses observations taken with the XMM-Newton telescope, 
an ESA science mission with instruments and contributions directly funded by ESA Member States and NASA.
MJP acknowledges support from the UK Science and Technology Facility Council (STFC) grant number ST/S000216/1.
MS would like to extend their gratitude towards Vladimir Yershov and 
Eduardo Ojero for their outstanding support with the XMM-OM software. 
MS would like to thank Michele Trenti
for sharing the source code for their cosmic variance calculator.
We thank the anonymous referee for their constructive report to further improve this manuscript.

\section{Data Availability}
\label{sec:9}
The data used in this article can be obtained from the \textit{XMM-Newton} Science archive (XSA) at \url{https://www.cosmos.esa.int/web/xmm-newton/xsa}. We provide 
the source list used in this paper as a supplementary table with the online version of the paper.
Other supporting material related to this article is available on a 
reasonable request to the corresponding author.

\bibliographystyle{mnras}
\bibliography{References}

\appendix
\section{Cross-correlating the Catalogues}
\label{sec:cross-corr}

\begin{figure*}
  \centering
  \hspace*{-0.1cm}\includegraphics [width=0.75\textwidth]{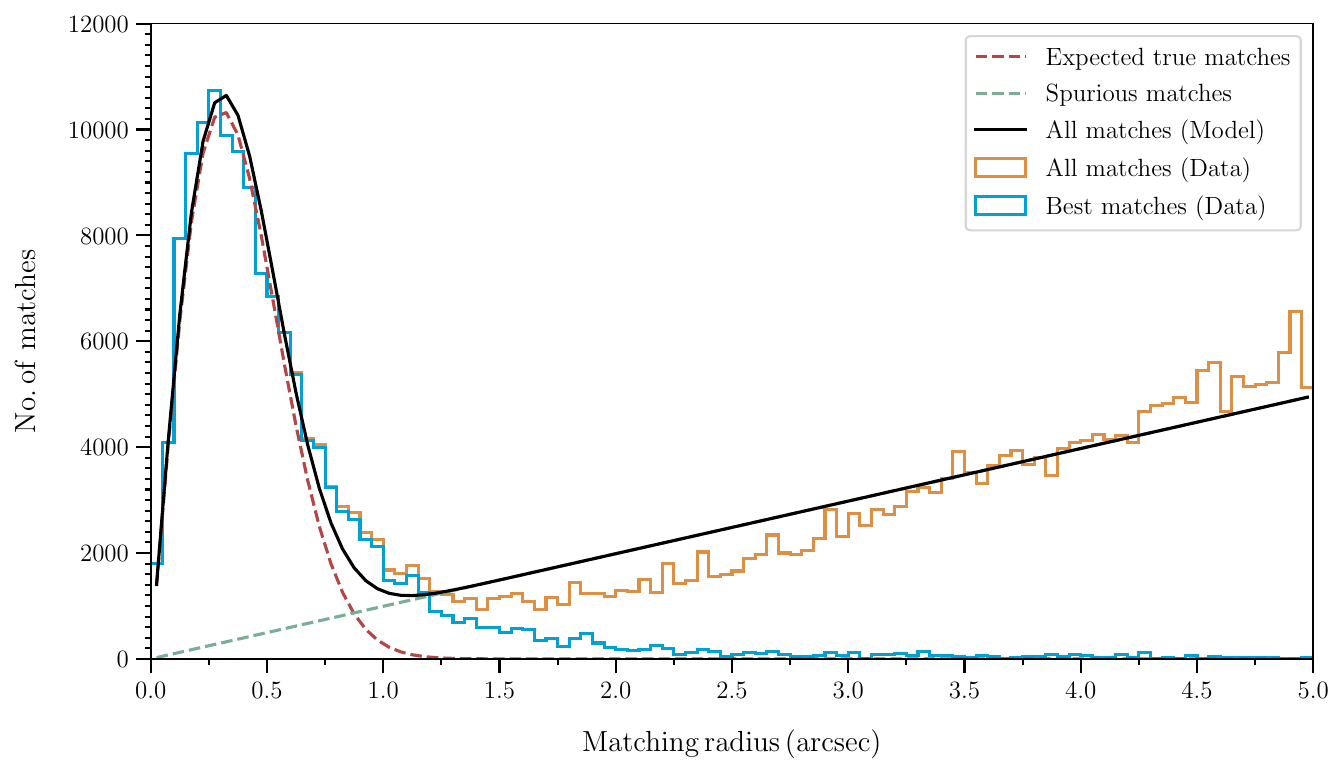}
  \caption{The angular offset distribution histograms are plotted in this figure.
  The yellow and blue histograms represent the angular separations corresponding to \textit{all} and \textit{best (closest)} matches between the UVW1 source list and the COSMOS2015 catalogue, within a given matching radius.
  The black solid line shows the expected composite model (Rayleigh + linear)
  of all matches fitted to the distribution of all matches. 
  The components of the composite model, i.e., the Rayleigh and the linear models, representing the true and spurious matches, are plotted in dashed red and green coloured lines respectively.}
  \label{fig:app_1}
\end{figure*}

\begin{figure}
    \centering
    \hspace*{-0.1cm}\includegraphics[width=0.95\columnwidth]{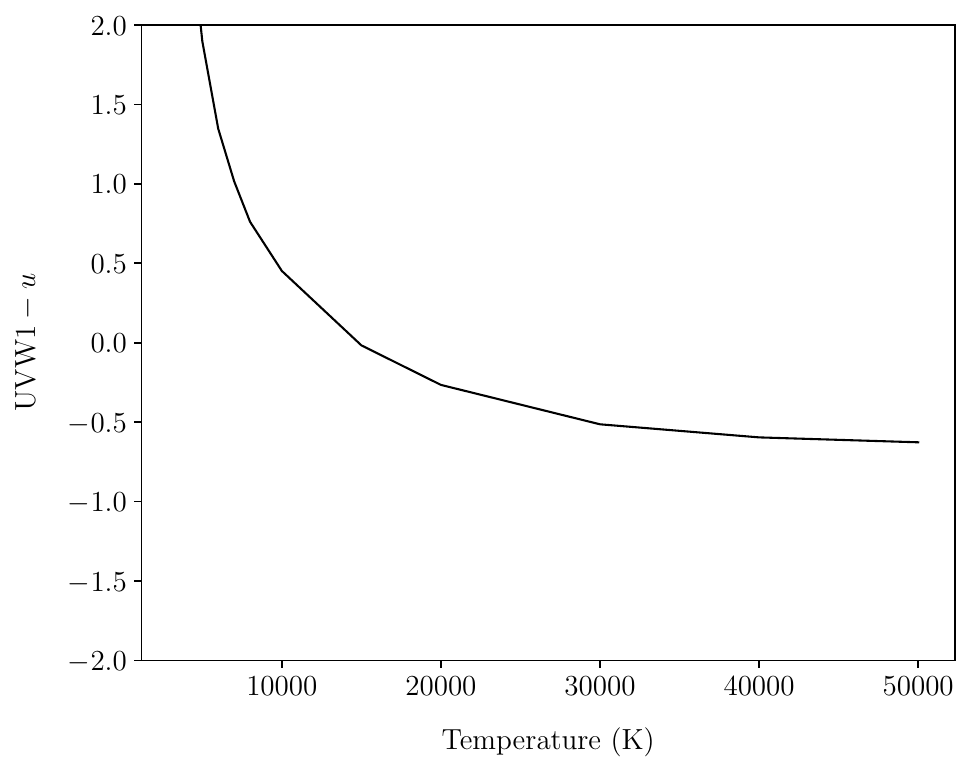}
    \caption{UVW1$-u$ colour as a function of the photospheric temperature of the stars.}
    \label{fig:app_2}
\end{figure}

\begin{figure}
    \centering
    \hspace*{-0.2cm}\includegraphics[width=0.95\columnwidth]{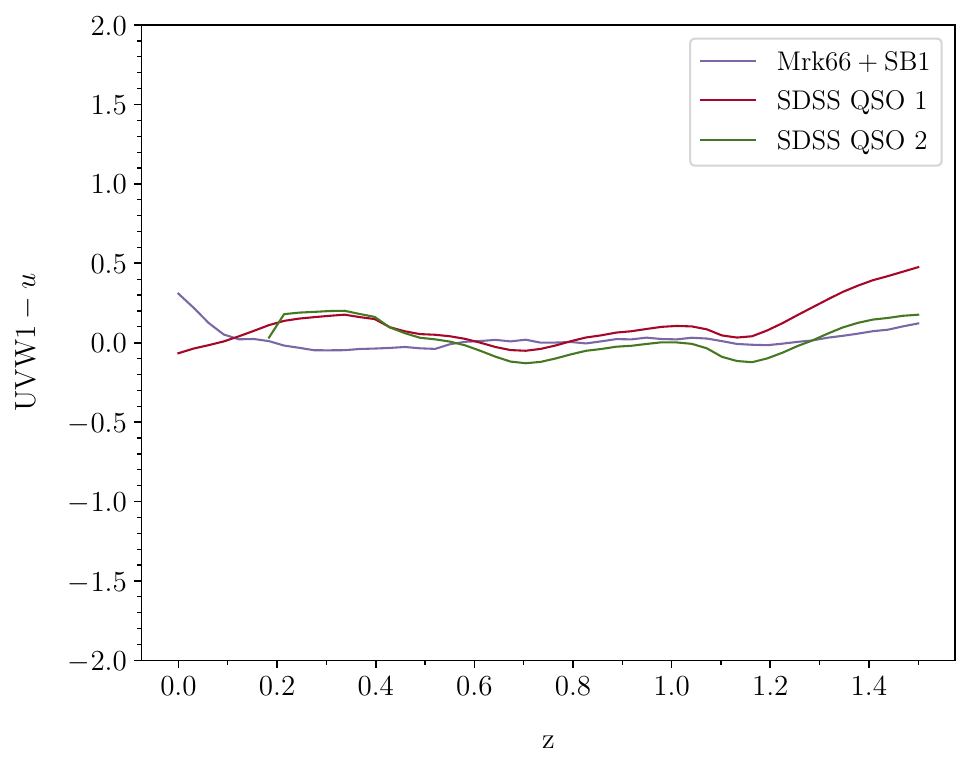}
    \caption{UVW1$-u$ colour tracks for the extended starburst (Mrk66+SB1)
    and the two SDSS average quasar templates as a function of 
    redshifts.}
    \label{fig:app_3}
\end{figure}

In this Section, we calculate the appropriate matching radius for cross-matching our
source list with the ancillary catalogues.

\subsection{Modeling the source-distribution}
The UVW1 source list is matched to the COSMOS2015 catalogue 
with a matching radius of 5 arcsecs.
We plot the distribution of the offsets in Fig. \ref{fig:app_1}.
The yellow and blue histograms represent the `all' and `best' matches
between the COSMOS2015 catalogue and our source list, as a function of matching radius.
Following \citetalias{2022MNRAS.511.4882S}, we fit a distribution of the form
\begin{equation}
  D(x) = \,
  A\, \frac{x}{\sigma^2} \,
  \mathrm{exp}\left({-\frac{x^2}{2\sigma^2}}\right) + m\,x,
  \label{eqn:spu}
\end{equation}
to the distribution of all matches, keeping $A,\,\sigma,\,$ and $m$ as 
free parameters, where $x$ is the distance in arcseconds between the UVW1 source position and COSMOS2015 counterpart position. The first part of the Equation
\ref{eqn:spu} represents the Rayleigh distribution predicted for the 
true shape of the actual counterparts \citep{2012MNRAS.426..903P}
and the second part shows a straight line for the distribution of 
the spurious counterparts, growing linearly with the matching radius.
The fit results are $5251 \pm 121$ sources, 
$0.308 \pm 0.005$ arcsec and $994 \pm 21$ sources per square arcsec
for $A,\,\sigma,\,$ and $m$, respectively.
Fig. \ref{fig:app_1} shows the two elements of the fit, the linear
distribution with a slope $m = 994$ sources per square arcsec 
in green dashed, and the Rayleigh
distribution with amplitude $A = 5251$ and width $\sigma = 0.308$ in 
dashed red curves, respectively. The solid black curve shows the total
model distribution for all matches.

Fig. \ref{fig:app_1} illustrates the distributions that can aid in approximating the optimal angular radius to minimise the occurrence of spurious matches. As observed in Fig. \ref{fig:app_1}, the modelled distribution of true matches (depicted by the red curve) decreases sharply beyond a matching radius of 1 arcsec, whereas the count of spurious matches continues to rise.
Hence, it appears that selecting a 1 arcsec radius strikes a suitable balance, ensuring an adequate number of matches for constructing the LF while effectively mitigating the presence of spurious sources.

To validate this choice, we must assess how many sources are spuriously matched to our UVW1 sources within a 1 arcsec radius. One approach is to determine the number of sources that fall under the linear distribution in Fig. \ref{fig:app_1} up to a specified matching radius.

For matching radii of 1 and 1.5 arcsecs, we find 504 (8 percent) and 1701 (25 percent) sources, respectively, that appear to be spurious. However, it is important to note that these numbers represent upper limits, particularly for smaller offset radii, where the linear distribution also includes contributions from the distribution of actual matches (i.e. those from the Rayleigh distribution).

In the subsequent Section, we will elaborate on additional measures employed to impose stricter constraints on the color characteristics of the sources. This is essential to maintain a reliable source list while extending the cross-matching radius to 1.5 arcseconds.

\subsection{UVW1\textit{-u} colour-cut}
\label{sec:colour_cut}

\begin{figure}
    \centering
    \includegraphics[width=0.95\columnwidth]{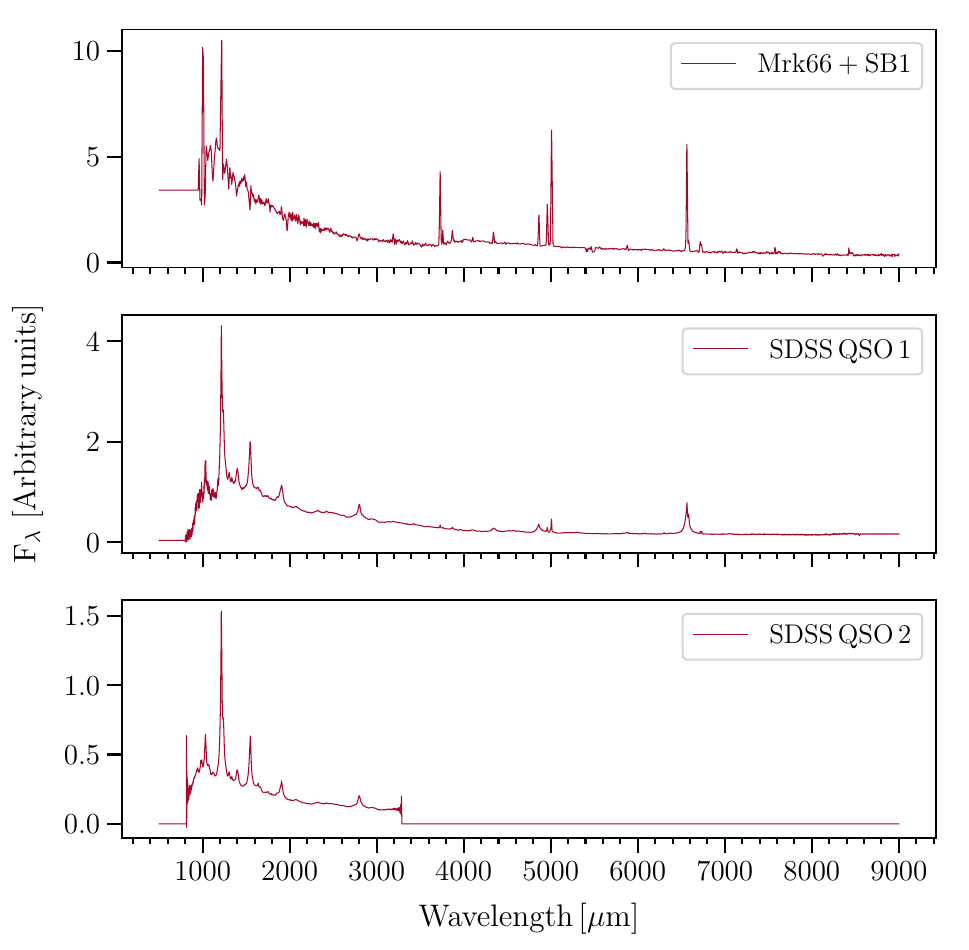}
    \caption{The spectral templates used to calculate the UVW1$-u$
    colours of the extended starburst galaxy (Mrk66+SB1) and the two average SDSS quasar templates.}
    \label{fig:app_4}
\end{figure}

\begin{figure}
    \centering
    \includegraphics[width=0.95\columnwidth]{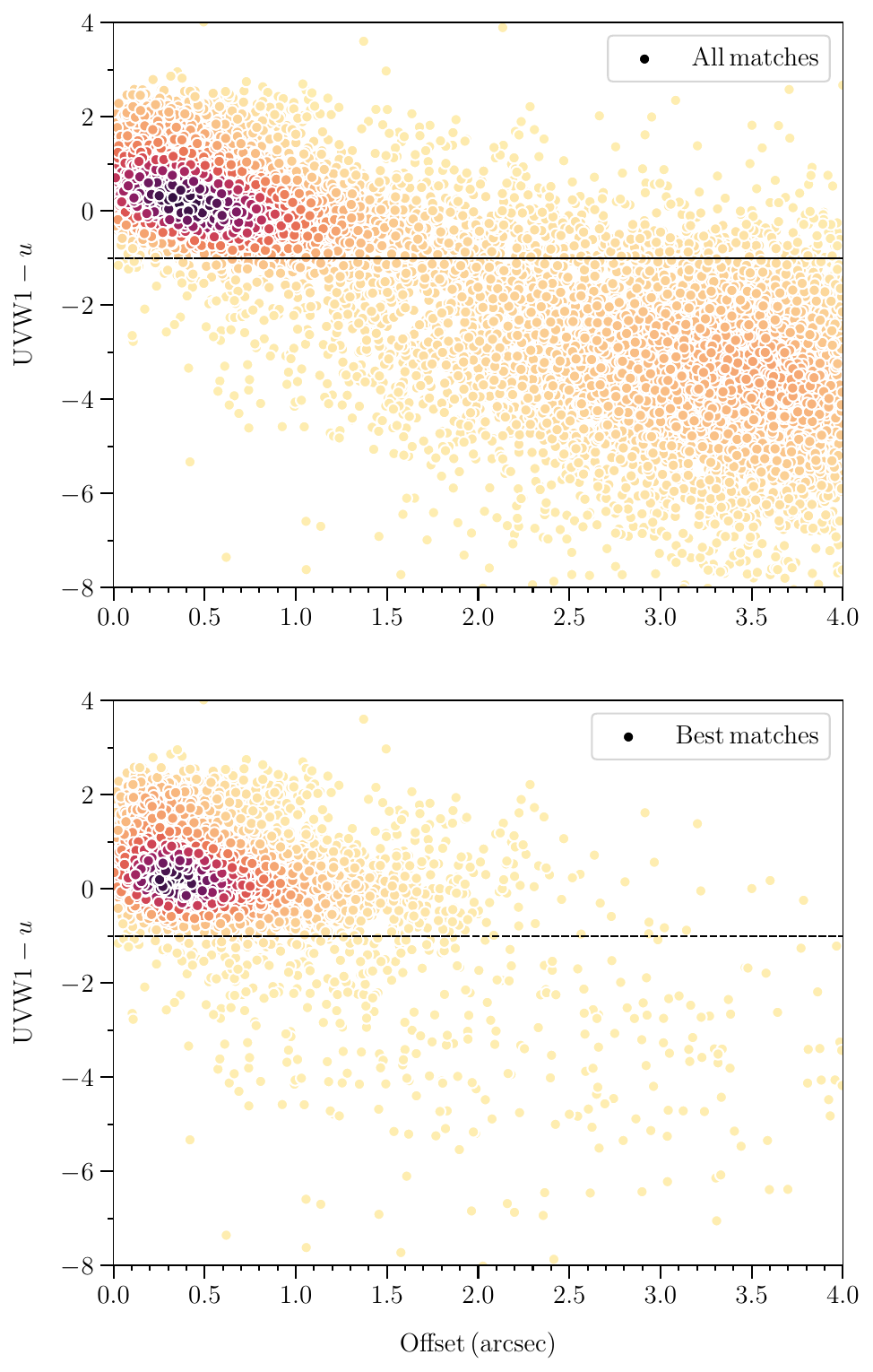}
    \caption{The UVW1$-u$ colours of the COSMOS UVW1 sources as a
    function of the matching radius for matches with the COSMOS2015 catalogue. 
    The density of the sources increases from
    lighter to darker shades. The top and bottom panels separately
    show the distributions for \textit{all} and the \textit{best (closest)} matches
    with the COSMOS2015 catalogue. The black dashed line at 
    UVW1$-u$ $= 1$ shows the colour-cut applied to our source list.}
    \label{fig:app_5}
\end{figure}

\begin{figure*}
  \centering
  \hspace*{-0.1cm}\includegraphics [width=0.75\textwidth]{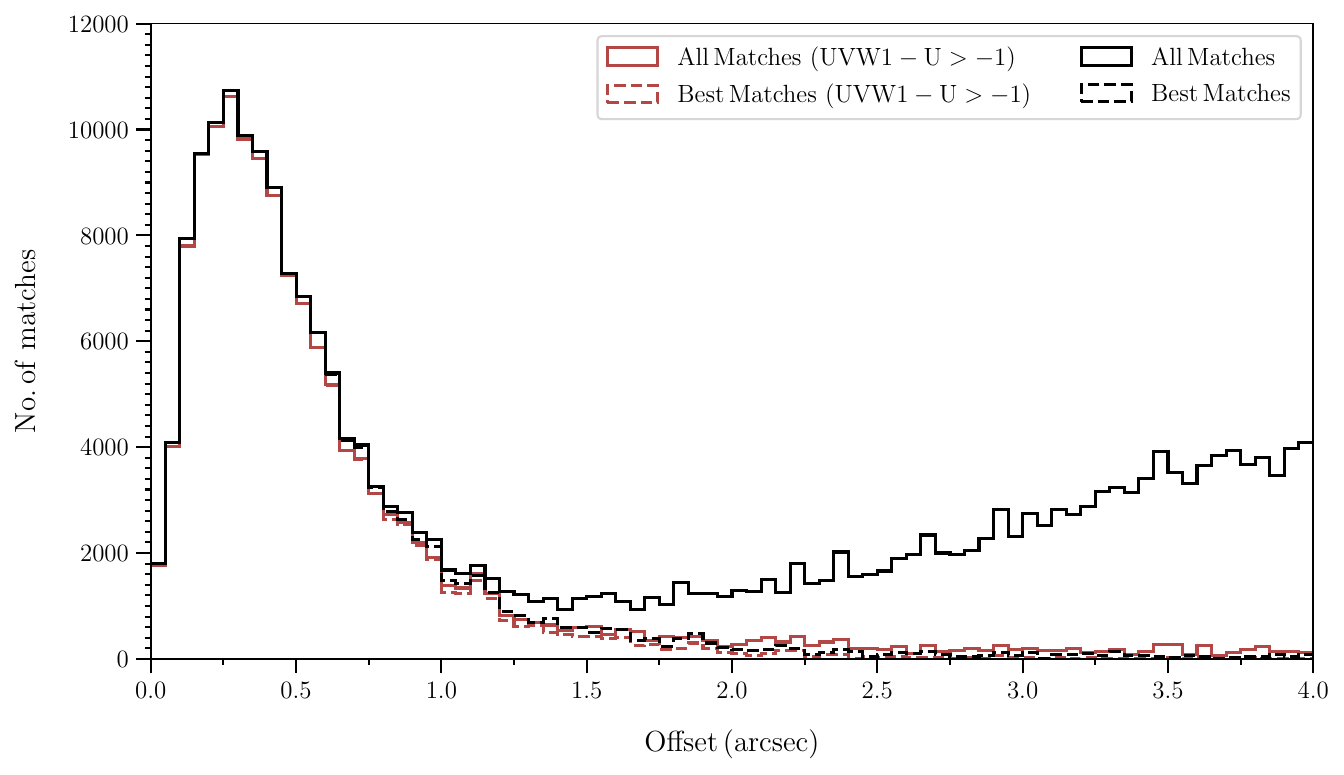}
  \caption{The distribution of matches from the COSMOS 2015 catalogue
  for our UVW1 sources as a function of offset radius. The solid black histogram represents all matches within a given matching radius. The dashed black histogram represents the best matches.
  Solid and dashed red histograms show the distributions of all and best matches after applying the colour-cut UVW1$-u$ $> -1$.}
  \label{fig:app_6}
\end{figure*}

To avoid missing out on genuine counterparts, we need to increase the matching radius while making sure that there are not a lot of spurious matches.
To achieve this, we introduce additional constraints on the colour properties of the sources. 
We examine how blue a source can physically be by calculating
the UVW1$-u$ colours for stars, quasars, and galaxies. We obtain the $u$ colours using CFHT $u$-band photometry from the COSMOS2015 catalogue.

For stars, we rely on synthesised stellar spectra derived from the ATLAS9 project \citep{2003IAUS..210P.A20C}, which incorporate stellar abundances from \citet{1998SSRv...85..161G}. We calculate the UVW1$-u$ colours for these spectra and plot their relationship with photospheric temperatures in Fig. \ref{fig:app_2}. As temperatures rise, the UVW1$-u$ colours shift toward a progressively bluer range, eventually stabilizing near a value of approximately $-0.5$ at the highest photospheric temperatures.

Our calculations for quasar colours are based on average SDSS spectral templates from  \citet{2001AJ....122..549V} and \citet{2016AJ....151..155H}. For galaxies, we adopt the starburst galaxy template (SB1) outlined in \cite{1996ApJ...467...38K}, extending it beyond 1250 Å using the spectrum of Mrk 66 from \citet{1998ApJ...495..698G}. In Fig. \ref{fig:app_3}, we present the UVW1$-u$ colour tracks for these galaxies.
We show the model spectra for the extended starburst galaxy and the quasar templates
(labelled SDSS QSO 1 and SDSS QSO 2 from \citet{2001AJ....122..549V} and \citet{2016AJ....151..155H}, respectively) in Fig. \ref{fig:app_4}.

Based on our analysis thus far, it is theoretically impossible for the UVW1$-u$ colours to be bluer than $-0.5$. It is important to note that we have not factored in the effects of reddening in these calculations. Consequently, we maintain a conservative colour threshold of UVW1$-u$ $> -1$ for our COSMOS UVW1 source list.
In Fig. \ref{fig:app_5}, the distribution of UVW1$-u$ colours for UVW1 sources are plotted against the offset radius for matching with the COSMOS2015 catalogue.

To assess the impact of this colour-cut, we have plotted the distribution of COSMOS 2015 counterparts for our UVW1 sources both before and after applying the colour-cut in Fig. \ref{fig:app_6}. After implementing the colour-cut, the distribution of best matches (shown as the solid red histogram) closely aligns with the distribution of best matches (shown as red and black dashed histograms).
This trend holds up to a 2.0 arcsec matching radius. Consequently, we can confidently use this 2.0 arcsec matching radius for cross-referencing the UVW1 source list with other ancillary catalogues, provided we maintain the UVW1$-u$ $> -1$ color-cut. 
However, for added caution, we opt for a conservative matching radius of 1.5 arcseconds in our cross-matching process.

\section{Spurious sources/detections in the UVW1 source list}
\label{sec:spurious_sources}

In this Section we assess the potential number of spurious sources that are likely to be present in our UVW1 source list and then proceed to estimate how many of these spurious sources might find their way into the catalogue we use to construct the LF.
We can use the number of UVW1 sources not matched to the background catalogue (COSMOS2015) sources to get an empirical estimate of the number of spurious UVW1 sources. We assume that the majority of spurious sources do not have a viable counterpart in the COSMOS2015 catalogue, whereas all the real sources should.

The first step is to estimate what fraction of spurious sources are likely to be matched to sources in the COSMOS2015 catalogue.
To do this, we employ a list of randomly generated source positions within the COSMOS UVW1 image, originally created for the completeness simulations (as detailed in Section \ref{sec:2.3}).
We cross-referenced the positions of these synthetic sources, totalling 58,131 in number, with the COSMOS2015 catalogue, resulting in 7,461 matches (equivalent to 12.8 percent of the synthetic sources).
Assuming that each synthetic source has a UVW1 AB magnitude of 23.02 (as estimated in Section \ref{sec:2.3} as the survey magnitude limit) and applying the UVW1$-u>-1$ colour-cut, the number of matches reduces to 1,199 (approximately 2.1 percent).

Turning now to our real UVW1 source list, applying the colour-cut described in 
Section \ref{sec:colour_cut} and the magnitude cut (UVW1 AB magnitude < 23.02), the number of UVW1 sources unmatched to COSMOS2015 is 568.
As we expect valid counterparts to be found for the vast majority of the real UVW1 sources, this number (568) serves as a reasonable estimate for the potential number of spurious sources within our UVW1 catalogue.
Based on our simulations, we know that only about 2.1 percent of spurious sources will successfully match with COSMOS2015 counterparts and be included in the catalogue we use for constructing the Luminosity Functions (LF).
This translates to an expected count of 12 out of 5,967 matches. Consequently, the contamination in the matched catalogue we use for the LF is likely to be around 0.2 percent.

\section{The variation of the dust extinction across the field}
\label{sec:dust_ext}

\begin{table}
  \setlength{\tabcolsep}{18pt}
  \centering
  \caption{Here we tabulate the coordinates of the XMM-OM sightlines for the centre of each subfield that constitute the COSMOS UVW1 image along with the dust extinction for those sightlines. The first two columns are pointing sky coordinates RA and DEC (in degrees) and the third column shows the $A_\mathrm{UVW1}$ values in the direction.}
    \begin{tabular}{ccc}
    \hline\hline
    \noalign{\vskip 0.5mm}
    RA &
    DEC  &  
    $A_\mathrm{UVW1}$  \\
    (deg) &
    (deg) &
    (mag) \\
    \hline
    \noalign{\vskip 0.5mm}
    10 02 26.41  &  +02 42 36.0 & 0.103 \\ 
    10 02 26.41  &  +02 27 36.0  & 0.097 \\
    10 02 26.41  &  +02 12 36.0  & 0.089 \\
    10 02 26.41  &  +01 57 36.0 & 0.097 \\
    10 02 26.41  &  +01 42 36.0  & 0.113 \\
    
    10 01 26.39  &  +02 42 36.0  & 0.101 \\
    10 01 26.39  &  +02 27 36.0  & 0.104 \\
    10 01 26.39  &  +02 12 36.0  & 0.083 \\
    10 01 26.39  &  +01 57 36.0  & 0.099 \\
    10 01 26.39  &  +01 42 36.0  & 0.100 \\
    
    10 00 26.38  &  +02 42 36.0 & 0.103 \\
    10 00 26.38  &  +02 27 36.0 & 0.092 \\
    10 00 26.38  &  +02 12 36.0  & 0.097 \\
    10 00 26.38  &  +01 57 36.0  & 0.093 \\
    10 00 26.38  &  +01 42 36.0  & 0.093 \\
    
    09 59 22.39  &  +02 42 36.0 & 0.101 \\
    09 59 26.39  &  +02 28 36.0  & 0.100 \\
    09 59 26.39  &  +02 12 36.0  & 0.106 \\
    09 59 26.39  &  +01 57 36.0  & 0.103 \\
    09 59 26.39  &  +01 42 36.0  & 0.103 \\
    
    09 58 26.40  &  +02 42 36.0  & 0.096 \\
    09 58 26.40  &  +02 27 36.0  & 0.107 \\
    09 58 26.40  &  +02 12 36.0  & 0.101 \\
    09 58 26.40  &  +01 57 36.0  & 0.102 \\
    09 58 26.40  &  +01 42 36.0  & 0.102 \\
    \hline
    \end{tabular}\\
  \label{tab:e_b_v}
\end{table}

Given the expansive sky coverage of the COSMOS field, it is possible that there are varying levels of dust content along different lines of sight within this region, potentially causing fluctuations in dust extinction affecting UV light. As a result of this differential dust extinction across the UVW1 COSMOS image, it may be necessary to apply individual dust corrections to the UVW1 magnitudes.
To evaluate the extent of this variability, we have calculated the Galactic extinction values toward the central positions of all 25 sub-fields that collectively constitute the COSMOS UVW1 image. We employed the same methodology described in Section \ref{sec:3.1} to obtain these values, and they are presented in Table \ref{tab:e_b_v}, showing the range of Galactic extinction across the COSMOS UVW1 image.
Our calculations indicate a mean dust extinction value of 0.099 magnitudes, with the deviation from this mean value across the 25 sub-fields remaining within a narrow range of 0.003 magnitudes around this mean value.

\bsp
\label{lastpage}
\end{document}